\documentclass{aastex62}
\pdfoutput=1
\usepackage{amsmath}
\usepackage{graphicx}
\usepackage{natbib}
\usepackage{appendix}

\newcommand\nnhp{N$_2$H$^+$}
\newcommand\dcop{DCO$^+$}

\shorttitle{N$_2$H$^+$ Probe}
\shortauthors{Qi et al.}

\begin{document}

\title{Probing CO and N$_2$ Snow Surfaces in Protoplanetary Disks with \nnhp{} Emission}

\correspondingauthor{Chunhua Qi}
\email{cqi@cfa.harvard.edu}

\author[0000-0001-8642-1786]{Chunhua Qi}
\affil{Harvard-Smithsonian Center for Astrophysics\\
60 Garden Street \\
Cambridge, MA 02138, USA}

\author{Karin I. \"Oberg}
\affiliation{Harvard-Smithsonian Center for Astrophysics\\
60 Garden Street \\
Cambridge, MA 02138, USA}

\author{Catherine C. Espaillat}
\affiliation{Department of Astronomy \& The Institute for Astrophysical Research\\
Boston University \\
725 Commonwealth Avenue \\
Boston, MA 02215, USA}

\author{Connor E. Robinson}
\affiliation{Department of Astronomy \& The Institute for Astrophysical Research\\
Boston University \\
725 Commonwealth Avenue \\
Boston, MA 02215, USA}

\author{Sean M. Andrews}
\affiliation{Harvard-Smithsonian Center for Astrophysics\\
60 Garden Street \\
Cambridge, MA 02138, USA}

\author{David J. Wilner}
\affiliation{Harvard-Smithsonian Center for Astrophysics\\
60 Garden Street \\
Cambridge, MA 02138, USA}

\author{Geoffrey A. Blake}
\affiliation{Division of Geological \& Planetary Sciences, MC 150-21 \\
California Institute of Technology \\
Pasadena, CA 91125, USA}

\author{Edwin A. Bergin}
\affiliation{Department of Astronomy \\
University of Michigan \\
500 Church Street
Ann Arbor, MI 48109, USA}

\author{L. Ilsedore Cleeves}
\affiliation{Department of Astronomy \\
University of Virginia \\
Charlottesville, VA 22903, USA}

\begin{abstract}
Snowlines of major volatiles regulate the gas and solid C/N/O ratios in the planet-forming midplanes of protoplanetary disks. Snow surfaces are the 2D extensions of snowlines in the outer disk regions, where radiative heating results in a decreasing temperature with disk height. CO and N$_2$ are two of the most abundant carriers of C, N and O.
\nnhp{} can be used to probe the snow surfaces of both molecules, 
because it is destroyed by CO and formed from N$_2$.
Here we present Atacama Large Millimeter/submillimeter Array (ALMA) 
observations of \nnhp{} at $\sim$0\farcs2--0\farcs4 resolution 
in the disks around LkCa~15, GM~Aur, DM~Tau, V4046~Sgr, AS~209, and 
IM~Lup. We find two distinctive emission morphologies: \nnhp{} is 
either present in a bright, narrow ring surrounded by extended 
tenuous emission, or in a broad ring.
These emission patterns can be explained by two different kinds of vertical temperature 
structures. Bright, narrow \nnhp{} rings 
 are expected in disks with a thick Vertically Isothermal Region above the Midplane (VIRaM) layer (LkCa~15, GM~Aur, DM~Tau)
where the \nnhp{} emission peaks between the CO and N$_2$ snowlines.  
Broad \nnhp{} rings come from disks with a thin VIRaM layer (V4046~Sgr, AS~209, IM~Lup). 
We use a simple model to extract the first sets of CO and N$_2$ snowline pairs and corresponding freeze-out temperatures towards the disks with a thick VIRaM layer. 
The results reveal a range of N$_2$ and CO snowline radii towards stars of similar spectral type, demonstrating the need for empirically determined snowlines in disks. 

\end{abstract}


\section{Introduction\label{sec:intro}} 

Snow lines, the transition regions in disk midplanes where volatiles such as 
H$_2$O, CO$_2$, and CO freeze out from the gas phase onto dust grains, may affect planet formation
through several different channels. Snow lines can locally enhance particle growth, 
boosting planetesimal formation efficiencies at specific radii through a combination of (1) increases in solid mass surface density exterior to snow line locations, (2) ``cold-head effects" that transport volatiles across snow lines, (3) pile-ups of dust just inside of the snow line in pressure traps, and (4) an increased ``stickiness" of icy grains at the H$_2$O snow line compared with bare particles \citep{Ciesla2006, 
Johansen2007, Chiang2010, Gundlach2011, Ros2013, Xu2017}. Snow lines also 
regulate the bulk elemental composition of planets and planetesimals, 
including the elemental C/N/O ratios \citep{Oberg2011b, Oberg2016, Piso2016}. 

Snowline locations can in principle be calculated using known disk midplane 
density and temperature structures, and volatile desorption and adsorption rates.
Desorption rates depend sensitively on volatile binding energies, however, and
laboratory experiments have demonstrated that CO and N$_2$ binding energies can vary by up to 50\%, depending on the structure and composition or water content of the ice they are adsorbed on \citep{Fayolle2016}. 
For the pressures present in disks, CO can therefore freeze out anywhere from $\sim$21 to 32~K, and N$_2$ between $\sim$18 and 27~K \citep{Collings2004,Oberg2005,Bisschop2006,Fayolle2016}.  Furthermore disk snowline locations are affected by the dynamics of the gas and dust. Pebble drift is especially relevant, since it can move snowlines closer to the star than would be expected from desorption-adsorption balance in static disks \citep{Piso2015}.
Accurate snowline predictions are therefore not possible, and 
we instead rely on observations of snowlines to benchmark disk models. 

Direct observations of snowlines through the identifications of sharp 
radial drop-offs in emission from the snowline volatile is challenging 
for CO and impossible for N$_2$. The latter prohibition is due to N$_2$'s 
lack of dipole moment, and therefore lack of observable rotational lines. 
Optically thin isotopologues of CO can be used to constrain the CO snowline 
location \citep{Qi2011,Schwarz2016,Huang2016,Zhang2017}, but it is difficult to distinguish 
between a decreasing gas surface density or CO abundance without independent
constraints on the former.
This issue is compounded by
vastly different disk temperature structures, which can make the expected CO emission drop at the CO snowline 
quite subtle (see discussions in Section 4.3).

An alternative approach to constrain snowline locations is to image the 
distribution of a chemically related species, whose abundance is connected
to the freeze-out of the desired volatile through a  simple set of
chemical reactions. \nnhp{} has been used to 
chemically image CO snowlines, exploiting the fact that \nnhp{} is quickly
destroyed by gas-phase CO and therefore only expected to be present in disks
exterior to the CO midplane snowline \citep{Qi2013c}.
The distribution of \nnhp{} also depends on the
N$_2$ snowline location since it forms through proton transfer
from H$_3^+$ to N$_2$. In cloud cores the depletion of \nnhp{} at low temperatures has indeed been successfully used to constrain N$_2$ freeze-out locations \citep{bergin2002}, but it has not yet been attempted in disks.

Interpretation of chemical images of snowlines is not always expected to be straightforward, however. \citet{vantHoff2017} modeled 
the relationship  between CO freeze-out and \nnhp{} emission and found that the \nnhp{} emission peaks  exterior to the CO snowline 
because of CO sublimation and \nnhp destruction in the warmer layers immediately above the midplane \citep[see Figure. 5 in ][]{vantHoff2017}. The disk height at which CO sublimates into the gas-phase above the CO midplane snowline depends on the steepness of the disk vertical temperature gradient.
The relationship between the \nnhp{} morphology and the CO snowline location, and therefore whether \nnhp{} imaging can be used to efficiently constrain the CO (and N$_2$) snowlines is then closely related to the disk vertical temperature structure.

Here we present Atacama Large 
Millimeter/submillimeter Array (ALMA) observations of \nnhp{} $3-2$ emission around 279.5~GHz
at $\sim$0\farcs2--0\farcs4 resolution in the disks around LkCa~15, 
GM~Aur, DM~Tau, V4046~Sgr, AS~209 and IM~Lup. These disks span a range of disk thermal vertical structure and therefore present a perfect test-bed for the utility of \nnhp{} in probing N$_2$ and CO snowlines in different kinds of disks. We
describe the sample and observations in Section 2, present the results and 
analysis in Section 3 and discussions in Section 4, and summarize 
our conclusions in Section 5. 

\section{Observations} \label{sec:obs}

\subsection{Sample selection}

\begin{deluxetable*}{lcccccccc}
\tablecaption{Stellar Properties \label{tab:stellar}}
\tablewidth{0pt}
\tablehead{
\colhead{Source\textsuperscript{a}} & \colhead{R.A.} & \colhead{Decl.} & \colhead{L$_{*}$} &
\colhead{Age} & \colhead{M$_{*}$\textsuperscript{b}} & \colhead{$\dot{M}$} &
\colhead{T$_{eff}$} & \colhead{Distance} \\
\colhead{} & \colhead{(J2000)} & \colhead{(J2000)} & \colhead{(L$_{\odot}$)} &
\colhead{(Myr)} & \colhead{(M$_{\odot}$)} & \colhead{(10$^{-9}$M$_\odot$ yr$^{-1}$)} &
\colhead{(K)} & \colhead{(pc)} 
}
\startdata
LkCa 15 & 04 39 17.80 & +22 21 03.1 & 1.0 & 2.0 & 1.2 & 6.3 &
4365 & 158 \\
GM Aur &  04 55 10.99 & +30 21 59.0 & 1.6 & 2.5 & 1.2 & 7.9 &
4786 & 159 \\ 
DM Tau & 04 33 48.75 & +18 10 09.7 & 0.24 & 4.0 & 0.52 & 4.0 & 3715 &
145 \\
V4046 Sgr & 18 14 10.48 & -32 47 35.4 & 0.49,0.33 & 24 & 1.75
& 0.5 & 4370 & 72 \\
AS 209 & 16 49 15.30 & -14 22 09.0 & 1.4 & 1.0 & 0.83 & 50 & 4266
& 121 \\
IM Lup & 15 56 09.2 & -37 56 6.5 & 2.6 & 0.5 & 0.89 & 13 & 4365 &
158 \\
\enddata

\tablecomments{\textsuperscript{a} The source
    information are from \citet{Andrews2018a} and the references within, except for V4046~Sgr from \citet{Huang2017} and the references within. \textsuperscript{b} Stellar masses for LkCa~15, GM~Aur and DM~Tau are fit, given the inclination determined from the continuum fitting.} 
\end{deluxetable*}
  
The disk sample consists of 6 T Tauri disks, which constitute a sub-set of the 12 disks observed in a multitude of molecular lines with the SMA as part of the DISCS survey \citep{Oberg2010, Oberg2011a}. 
Six of the DISCS targets were detected in N$_2$H$^+$ by the SMA and these constitute our sample.
It consists of four T Tauri stars with large central cavities in millimeter dust emission (LkCa~15, GM~Aur, DM~Tau,
V4046~Sgr) and two T Tauri stars without such cavities (AS~209 and IM~Lup). Table~1 
shows the detailed stellar properties for the 6 sources.  

\subsection{Observation setup}

\begin{deluxetable*}{lccccccc}
\tablecaption{ALMA observation details\label{tab:obs}}
\tablewidth{0pt}
\tablehead{
\colhead{Source} & \colhead{Date} & \colhead{Antennas} & \colhead{Baselines} &
\colhead{On-source} & \colhead{Bandpass} & \colhead{Phase} &
\colhead{Flux\textsuperscript{b}} \\
\colhead{} & \colhead{} & \colhead{} & \colhead{(m)} &
\colhead{integration (min)} & \colhead{Calibrator} & \colhead{Calibrator} & 
\colhead{Calibrator}
}
\startdata
LkCa15/GM Aur\textsuperscript{a} & 2016 July 27 & 45 & 15-1100 & 19/17 & J0510+1800 & J0433+2905 & J0510+1800 (1.443 Jy)\\
              & 2016 August 16 & 42 & 15-1500 & 19/17 & J0510+1800 & J0433+2905 & J0510+1800 (1.554 Jy) \\
              & 2016 August 22 & 40 & 15-1500 & 19/17 & J0510+1800 & J0433+2905 & J0423-0120 (0.442 Jy) \\
DM Tau        & 2016 August 31 & 39 & 15-1800 & 53    & J0510+1800 & J0431+2037 & J0238+1636 (0.961 Jy) \\
V4046 Sgr     & 2016 April 30 & 41 & 16-640 & 20 & J1924-2914 & J1802-3940 & Titan \\
AS 209        & 2016 June 09 & 38 & 16-783 & 64 & J1517-2422 & J1733-1304 & J1733-1304 (1.618 Jy) \\
              & 2016 August 26 & 42 & 15-1500 & 32 & J1517-2422 & J1733-1304 & J1733-1304 (1.555 Jy) \\
IM Lup        & 2016 June 09 & 38 & 16-783 & 29 & J1517-2422 & J1604-4441 & J1517-2422 (2.265 Jy) \\
\enddata

\tablecomments{\textsuperscript{a} LkCa~15 and GM~Aur
were observed in the same scheduling blocks.\textsuperscript{b} Quasar flux averaged over all the spectral windows}
\end{deluxetable*}

The six disks were observed during ALMA Cycle 3 (project code
[ADS/JAO.ALMA\#2015.1.00678.S]) in Band 7 with a single spectral set-up consisting of 6 spectral windows with widths and resolutions ranging from 117.2 to 937.5 MHz, and 122.1 to 976.6 khz respectively. The \nnhp{} $3-2$ line was place in a 117.2 MHz wide spectral window with a native resolution of 122.1~khz (0.13 km~s$^{-1}$). Additional windows covered the 
\dcop{} 4--3 and H$_2$CO 6$_{0,6}$--5$_{0,5}$ lines. The \dcop{}  data will be presented in a future paper, and the H$_2$CO data is presented in Pegues et al. (subm.).  V4046~Sgr data were also presented in \citet{Kastner2018}.
The array configuration and calibrators for each observation are described in Table~\ref{tab:obs}.  

\subsection{Data reduction}

ALMA/NAASC staff performed the standard pipeline calibration tasks. 
Subsequent data reduction and imaging were completed with
CASA 4.7.0 \citep{McMullin2007}. For each disk, the data were separated with the spectral
windows in the upper and lower sideband. For each sideband, the
corresponding continuum was phase self-calibrated by combining the
line-free channels of the spectral windows and then imaged by CLEANing
with a robust parameter of 0.5. The self-calibrated tables were subsequently
applied to the spectral windows for each sideband, and the 284 GHz
continuum was obtained by combining the linefree channels of both sidebands. Table~\ref{tab:cont} 
lists the fluxes, rms values, beam sizes and position angles, as well as the disk inclinations and position angles. The latter two are based on Gaussian fits to the disk continuum: we determine the position angle of the 
disk major axis (measured east of north) and the disk inclination 
(0 is face-on) by fitting an elliptical Gaussian component on 
the high SNR continuum image using the CASA task IMFIT. 
The values for AS~209 and IM~Lup are consistent with those in \citet{Huang2018b} derived from deeper ALMA observations with much higher angular resolution.
The radius of the millimeter dust continuum was 
estimated by the size of the 3$\sigma$ contour of the image in the direction of the disk major axis.

\begin{deluxetable*}{lcclccc}
\tablecaption{Properties of the 284 GHz dust continuum and derived disk properties\label{tab:cont}}
\tablewidth{0pt}
\tablehead{
\colhead{Source} & \colhead{F$_{cont}$\textsuperscript{a}} 
& \colhead{Peak flux density\textsuperscript{a}} &
\colhead{Beam (P.A.)} & \colhead{Disk Incl.} & 
\colhead{Disk P.A.} &
\colhead{Radius ($3\sigma$} \\
\colhead{} & \colhead{(mJy)} & \colhead{(mJy beam$^{-1}$)} 
& \colhead{} & \colhead{(deg)} & \colhead{(deg)} &
\colhead{(au)}
}
\startdata
LkCa 15 & 281.0$\pm$6.9 & 13.01$\pm$0.31 & $0\farcs.25 \times 0\farcs22$ (-18$^\circ$.2)& 50.5$^{+2.3}_{-2.4}$ & 61.4$\pm$2.1 & 230 \\
GM Aur & 286.4$\pm$2.6 & 22.27$\pm$0.19 & $0\farcs29 \times 0\farcs18$ (-1$^\circ$.6)& 51.9$^{+0.8}_{-0.9}$ & 56.5$\pm$0.8 & 300 \\
DM Tau & 99.1$\pm$1.4 & 12.38$\pm$0.15 & $0\farcs21 \times 0\farcs18$ (-177$^\circ$.9)& 36.3$^{+2.3}_{-2.4}$ & 157.6$\pm$2.9 & 240 \\
V4046 Sgr & 576$\pm$12 & 49.58$\pm$0.97 & $0\farcs38 \times 0\farcs29$ (-73$^\circ$.1)& 33.6$^{+2.8}_{-4.0}$ & 74.5$\pm$5.7 & 110 \\ 
AS 209 & 289.0$\pm$4.5 & 40.45$\pm$0.56 & $0\farcs32 \times 0\farcs21$ (-70$^\circ$.9)& 35.0$^{+2.8}_{-3.1}$ & 90.5$\pm$4.1 & 180 \\
IM Lup & 243.7$\pm$6.2 & 54.1$\pm$1.1 & $0\farcs40 \times 0\farcs33$ (82$^\circ$.6)& 44.2$^{+3.5}_{-3.9}$ & 139.9$\pm$5.0 & 360 \\
\enddata
\tablecomments{\textsuperscript{a}Uncertainties do
    not include 10\% systematic flux uncertainties.}
\end{deluxetable*}

After self-calibration, the continuum were subtracted in the $uv$-plane
from the spectral windows containing \nnhp{}. Then the data
were binned at 0.15 km s$^{-1}$ and output to UVFITS file. Then UVFITS files
were loaded into MIRIAD \citep{Sault1995} for imaging. All \nnhp{} line observations were CLEANed with a Briggs
parameter of 2.0 for optimal sensitivity. 
 
\section{Results and Analysis} \label{sec:results}

\begin{deluxetable*}{lccccl}
\tablecaption{Line Observations \label{tab:line}}
\tablewidth{0pt}
\tablehead{
\colhead{Source} & \colhead{Integration range} &
\colhead{Channel rms} & \colhead{Mask axis} & \colhead{Integrated flux} &
\colhead{Beam (P.A.)} \\
\colhead{} & \colhead{(km s$^{-1}$)} & \colhead{(mJy beam$^{-1}$)} & \colhead{($''$)} &
\colhead{(mJy km s$^{-1}$)} & \colhead{}
}
\decimalcolnumbers
\decimals
\startdata
LkCa 15 & 2.7--10.2 &3.2& 6 & 1820 $\pm$ 23 & $0\farcs32 \times 0\farcs29$ (-23$^\circ$.0) \\
GM Aur  & 1.6--9.4 &3.2& 6 & 1487 $\pm$ 23 & $0\farcs36 \times 0\farcs24$ (-3$^\circ$.0) \\
DM Tau  & 4.5--8.0 &4.1& 6 & 1286 $\pm$ 40 & $0\farcs30 \times 0\farcs28$ (8$^\circ$.1) \\ 
V4046 Sgr & -0.9--7.2 &5.4 & 8 & 3761 $\pm$ 81 & $0\farcs59 \times 0\farcs49$ (-78$^\circ$.5) \\
AS 209 & 1.9--7.9 &2.7& 4 & 917 $\pm$ 15 & $0\farcs42 \times 0\farcs27$ (-70$^\circ$.3) \\
IM Lup & 1.8--7.5 &4.5& 6 & 2040 $\pm$ 49 & $0\farcs49 \times 0\farcs41$ (88$^\circ$.1) 
\enddata
\tablecomments{Column descriptions: (1) Source name. (2) Velocity range integrated across to calculate integrated flux. (3) Channel rms for bin size of 0.15 km s$^{-1}$. (4) Major axes of elliptical spectral extraction masks. (5) Integrated flux. Uncertainties do not include systematic flux uncertainties. (6) Synthesized beam dimensions}
\end{deluxetable*}

\subsection{Observational results}

\begin{figure}
\plotone{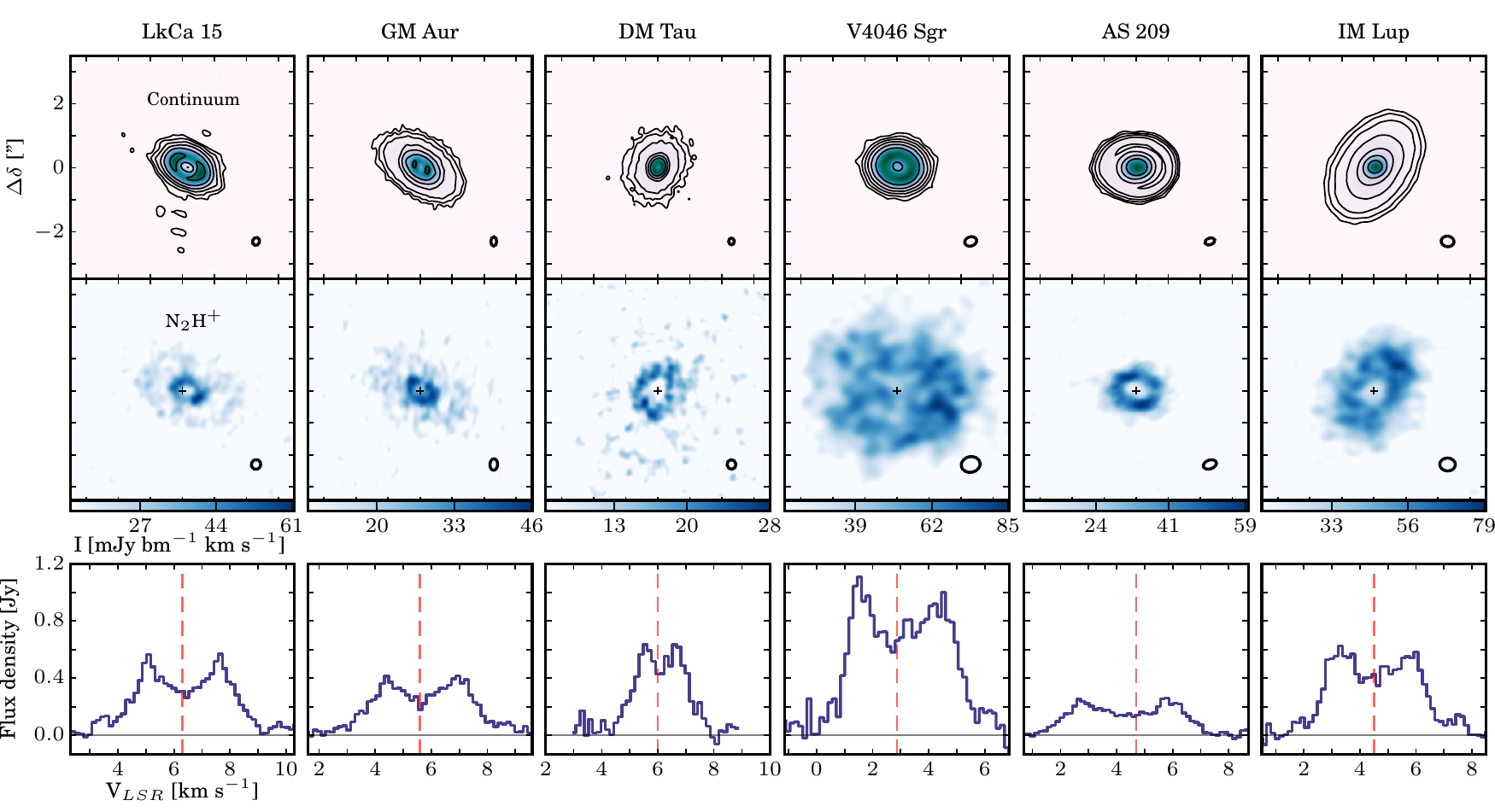}
\caption{Continuum images and \nnhp{} $3-2$ integrated intensity maps and spectra. First row: intensity maps of the 284 GHz dust continuum. Contours are drawn at [5, 10, 20, 40, 80, 160...]$\sigma$, where $\sigma$ is the rms listed in Table~\ref{tab:cont}. Note the continuum sub-structure in the LkCa~15, GM~Aur, DM~Tau and AS~209 disks.
Second row: integrated intensity maps of \nnhp{} $3-2$. Color bars start at 2$\sigma$, where $\sigma$ is the rms of the integrated intensity map. Synthesized beams are drawn in the lower right corners of each panel. The centroid of the continuum image is marked with a cross. Offset from the centroid in arcseconds is marked on the y axis of the upper left corner panel.
The \nnhp{} maps display a range of morphologies from narrow to broad rings. 
Third row: Spectra of \nnhp{} $3-2$ show the double-peaked shape characteristic of a Keplerian rotation disk. Even though there are 29 hyper-fine components for the \nnhp{} $3-2$ line (see Appendix), the dominant components spread less than 0.4 km~s$^{-1}$. The vertical red line marks the systemic velocity.}
\label{fig:mapspec}
\end{figure}

Figure~\ref{fig:mapspec} shows the 284 GHz continuum and the integrated intensity 
maps and spectra of \nnhp{} $3-2$ towards the six disks. The central dust cavities reported in LkCa~15, GM~Aur, DM~Tau, V4046~Sgr 
\citep{Hughes2009, Rosenfeld2013, Andrews2011, Huang2017} are resolved in 
the 284 GHz continuum. The multi-ringed dust structure resolved in the 
continuum of AS~209 is also consistent with those shown in \citet{Huang2017, Fedele2018, Guzman2018}. 

\begin{figure}
\includegraphics[width=\textwidth]{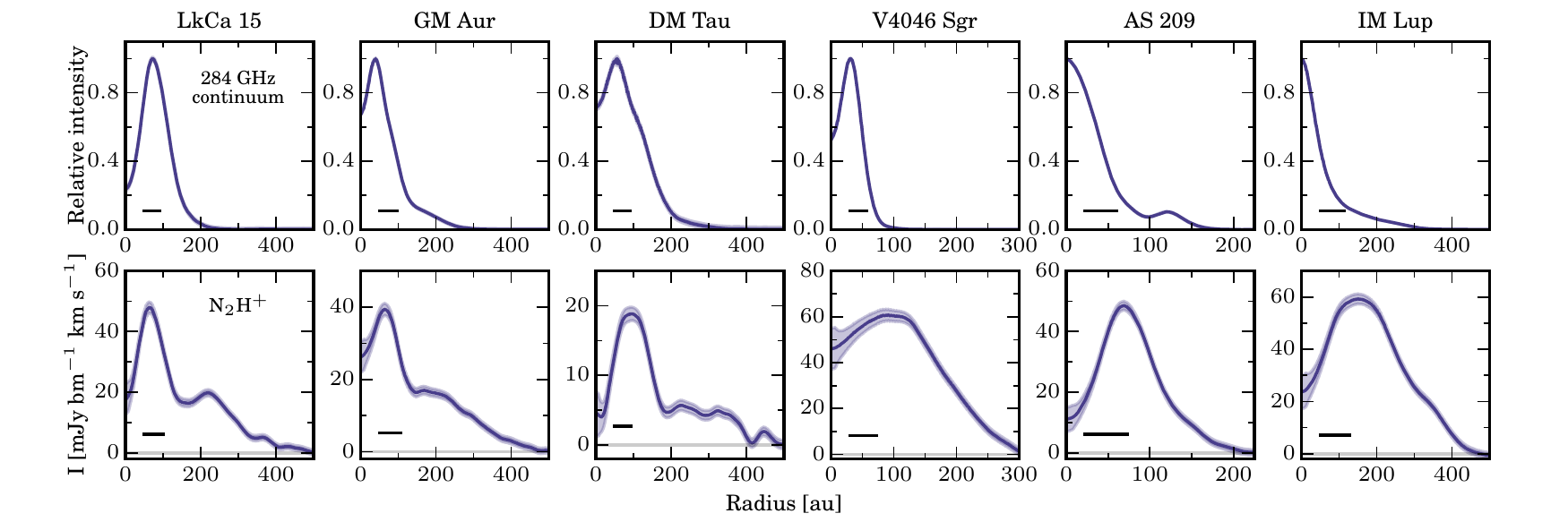}
\caption{Deprojected and azimuthally averaged 284 GHz continuum intensity and integrated \nnhp{} $3-2$ line intensity radial profiles. Top row: continuum intensity profiles normalized to peak values. Bottom row: radial profiles of integrated line intensity. The color shades show the standard deviation in pixel intensities calculated at each annulus. Adopted distances are listed in Table~\ref{tab:stellar}, and adopted position angles and inclinations are listed in Table~\ref{tab:cont}. Black bars in each panel represent the restoring beam major axes. 
Note the distinct N$_2$H$^+$ morphologies in the three leftmost disks -- a narrow ring followed by a low-level plateau, and the three rightmost disks -- a broad torus more extended than the continuum disk. \label{fig:profiles}}
\end{figure}

The \nnhp{} integrated intensity maps (second row in Figure~\ref{fig:mapspec}) were produced by summing over
channels in velocity ranges listed in Table~\ref{tab:line}, corresponding to where \nnhp{} emission was detected above the 2-3$\sigma$ level. The \nnhp{} spectra (third row in Figure~\ref{fig:mapspec}) were extracted from the image cubes (shown in the Appendix) using elliptical regions with centroid of the continuum image and shapes and orientations based on the inclination and position angle for each disk (Table~\ref{tab:cont}). The major axis of the elliptical region (Table~\ref{tab:line}) is chosen to cover the $>3\sigma$ line emission in the image cubes. The uncertainties are treated the same way as described in \citet[Section 3.2]{Huang2017}. Finally, Figure.~\ref{fig:profiles} 
shows the corresponding deprojected and azimuthally averaged radial profiles for each disk,
assuming the position angles and inclinations listed in Table~\ref{tab:cont}. 

 Based on the integrated intensity maps and radial profiles, the morphologies of the \nnhp{} emission can be divided
into two distinct groups: (1) a bright, narrow ring surrounded by tenuous extended
emission in the disks of LkCa~15, GM~Aur, and DM~Tau; (2) a broad torus extending across the dust disk for V4046~Sgr, AS~209, and IM~Lup. At first glance, AS~209 seems to straddle these two categories, but its \nnhp{} emission is ``broad'' compared to its compact continuum, placing it in the second group. 

A second, tenuous ring of
\nnhp{} can be seen in the disks of LkCa~15 (at $\sim$220 au), GM~Aur ($\sim$200 au), and IM~Lup
($\sim$350 au) in the radial profiles (and also in the channel maps). In all sources the second ring appears to 
be close to the edge of the mm dust continuum emission (Table~\ref{tab:cont}). In the case of IM~Lup 
and LkCa~15, the second \nnhp{} ring resides somewhat exterior to the second rings of  
H$^{13}$CO$^+$ and DCO$^+$ in the same disk:\footnote{Scaled with the updated distances.} 310 au for IM~Lup; 200 au for LkCa~15 \citep{Huang2017}, possibly related to CO desorption at these large radii.
The spatial coincidence between this second \nnhp{} ring and the continuum edge suggests that its appearance is due to an increased radiation penetration towards the disk midplane beyond the pebble edge. Enhanced radiation can cause a thermal inversion \citep{Cleeves2016}, resulting in N$_2$ and CO desorption, and should also increase the ionization level \citep{Bergin2016}, which may promote \nnhp{} production. The displacement between the second ring of \nnhp{} and that of H$^{13}$CO$^+$ and DCO$^+$ indicates a possible thermal inversion near or beyond the pebble edge.

\subsection{Effects of vertical temperature profiles on \nnhp{} emission }

In this paper, we argue that the morphology differences of \nnhp{} emission depend mainly on the disk vertical temperature structure. 
In the outer disk (beyond 10 au), the disk radial-vertical temperature structure is generally set by 
the radiation heating and dust growth/settling.
The temperature decreases deeper in the disk since the stellar radiation is
deposited in the atmosphere and reprocessed to lower energy photons that warm the disk interior. Near the midplane (z=0) where the disk is optically thin to its own radiation,
the temperature is approximately constant with height, maintained by the reprocessed
flux reaching to this layer, resulting in a Vertically Isothermal Region above the Midplane (VIRaM) layer.
The height of this layer can vary substantially from disk to disk;
the isothermal layer extends higher for disks that have more small grains that trap radiation at surface, whereas the isothermal layer  is lower for more settled disks (i.e., large dust grains are more concentrated towards the midplane \citep[see Figure. 5 in ][]{dAlessio2006}).
Above the VIRaM exterior to the CO midplane snowline, there is a vertical layer where N$_2$ is in the gas-phase, while CO is not.

\begin{figure}
\includegraphics[width=\textwidth]{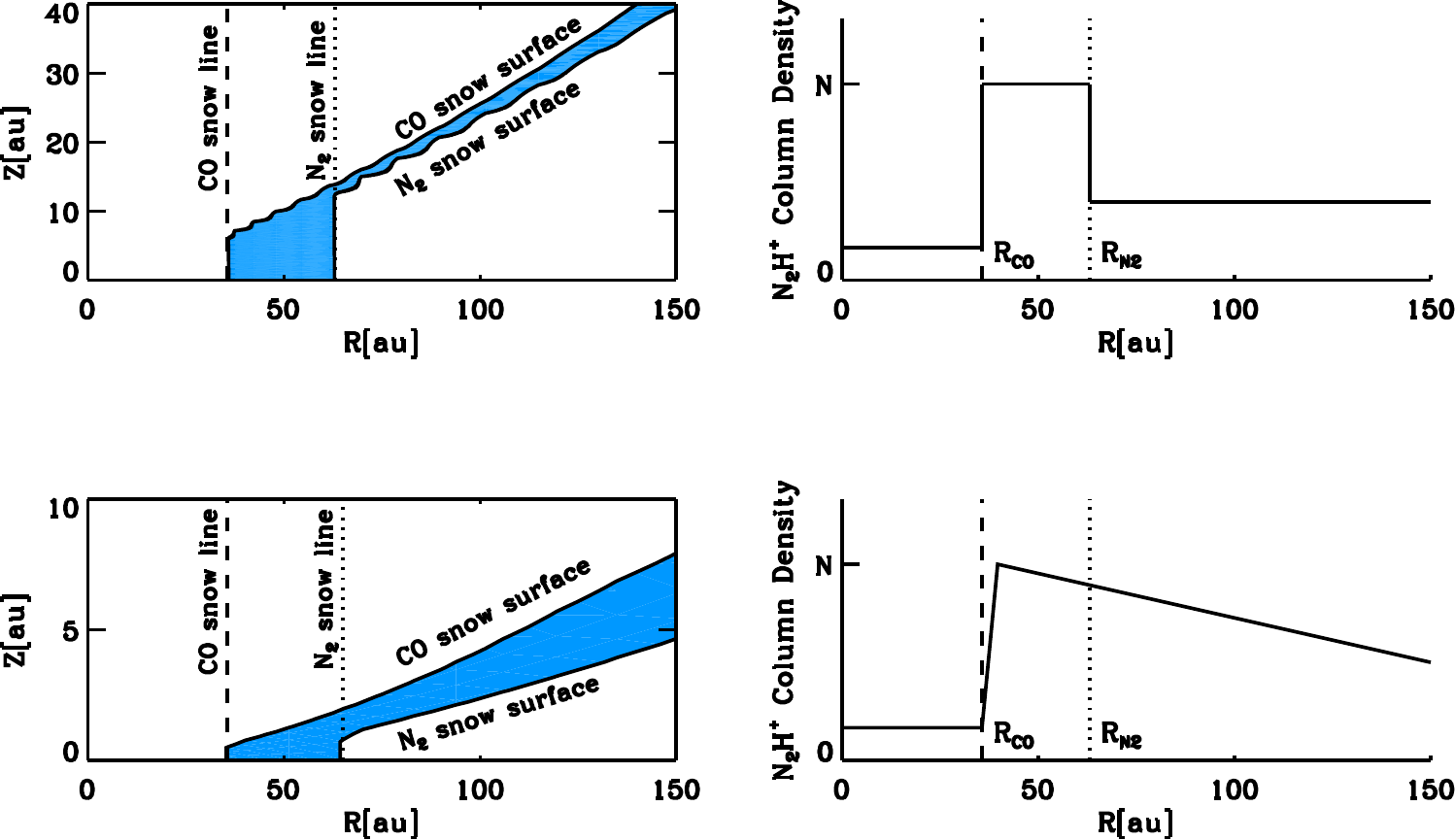}
\caption{\label{fig:n2hpsnow} Schematic representation of the effects of vertical temperature
structures on the \nnhp{} emission. {\em Left panels}: The snow surfaces of CO and N$_2$ in the disks with same midplane temperatures but different vertical temperature structures; in the upper panel the VIRaM layer extends through a substantial part of the disk, while in the lower panel the VIRaM layer is thin, and instead there is a vertical temperature gradient from the midplane to the disk surface. The solid blue shows the disk regions where N$_2$ is in the gas-phase and CO is frozen out, and thus where N$_2$H$^+$ should be abundant. {\em Right panels}: The expected corresponding \nnhp{} column density distributions. The thick VIRaM layer case results in sharp edges at the CO and N$_2$ midplane snowline, while the opposite is true for disks with a thin VIRaM layer.}
\end{figure}

We expect \nnhp{} to trace the CO and N$_2$ snow surfaces -- the 2D iso-temperature contours that correspond to freeze-out temperature of CO and N$_2$ across the disk. Figure~\ref{fig:n2hpsnow} visualizes the impact of different disk vertical temperature structures on the \nnhp{} radial column density profile. For disks with a
thick VIRaM layer (upper panels), the CO and N$_2$ snow surfaces extend vertically from
the midplane snowlines and we expect the 
\nnhp{} emission to present a
 bright, and narrow ring, with inner and outer radii constrained by the two snowlines. 
Beyond the \nnhp{} midplane snowline we expect some tenuous emission originating from higher disk layers where \nnhp{} exists in between the CO and N$_2$ snow surfaces. Because of a steep temperature gradient at these disk heights, and a comparably small difference in freeze-out temperature between CO and N$_2$, this contribution should be small and readily distinguishable from the emission component marking the CO and N$_2$ snowlines.

The lower panels of Figure.~\ref{fig:n2hpsnow} show that for disks
with a thin VIRaM layer the \nnhp{} emission profile traces the CO midplane snowline more indirectly and the N$_2$ snowline not at all. In such disks, snow surfaces are non-vertical resulting in an offset between the CO snowline and the inner edge of \nnhp{} emission, as described by \citet{vantHoff2017}. 
The connection between the \nnhp{} emission and the N$_2$ snowline is completely washed out because the vertical layer where \nnhp{} can exist 
above the CO snow surface is thick compared to the VIRaM layer.
Observationally disks of this kind should present broad \nnhp{} rings with inner edges slightly beyond the CO snowline and outer edges limited by total disk column density fall-offs in the outer disks.

\subsection{Disk vertical temperature structure models}
The two groups of observed \nnhp{} emission morphologies 
resemble those expected from disks with a thick or thin VIRaM layer 
(Figure.~\ref{fig:n2hpsnow}). To test whether the observed disks do fall into the two groups, and to retrieve snowline locations in disks with 
a thick VIRaM layer, we
use the D'Alessio Irradiated Accretion Disk (DIAD)
code \citep{dAlessio1998, dAlessio1999, dAlessio2001, dAlessio2005, dAlessio2006}
to model the disks, with special focus on the disk vertical temperature structures.  

DIAD considers 
a flared irradiated accretion disk in
hydrostatic equilibrium and the radial and vertical structure of the disks
are calculated self-consistently. For a given mass accretion rate ($\dot{M}$),
and viscosity coefficient ($\alpha$), the density and temperature
structure of this model is determined as described
by \citet{dAlessio1998, dAlessio1999}. We consider heating from the
mechanical work of viscous dissipation (relevant only in the midplane
of the inner disk), accretion shocks at the stellar surface, and
passive stellar irradiation, and follow the radiative transfer of that
energy with 1+1D calculations using the Eddington approximation and a
set of mean dust opacities (gas opacities are considered
negligible). The dust is  assumed to be a mixture of segregated
spheres composed of ``astronomical" silicates and graphite, with
abundances (relative to the total gas mass) of  $\zeta_{\rm sil} =
0.004$ and $\zeta_{\rm gra} = 0.0025$ \citep{Draine1984}: the
``reference" dust-to-gas mass ratio is $\zeta_{\rm ref} = 0.0065$.  At
any  given location in the disk, the grain size ($a$) distribution of
these dust particles is assumed to be a power-law, $n(a) \propto
a^{-3.5}$, between $a_{\rm min} = 0.005$\,$\mu$m and a specified $a_{\rm max}$.

\begin{figure}
\gridline{\fig{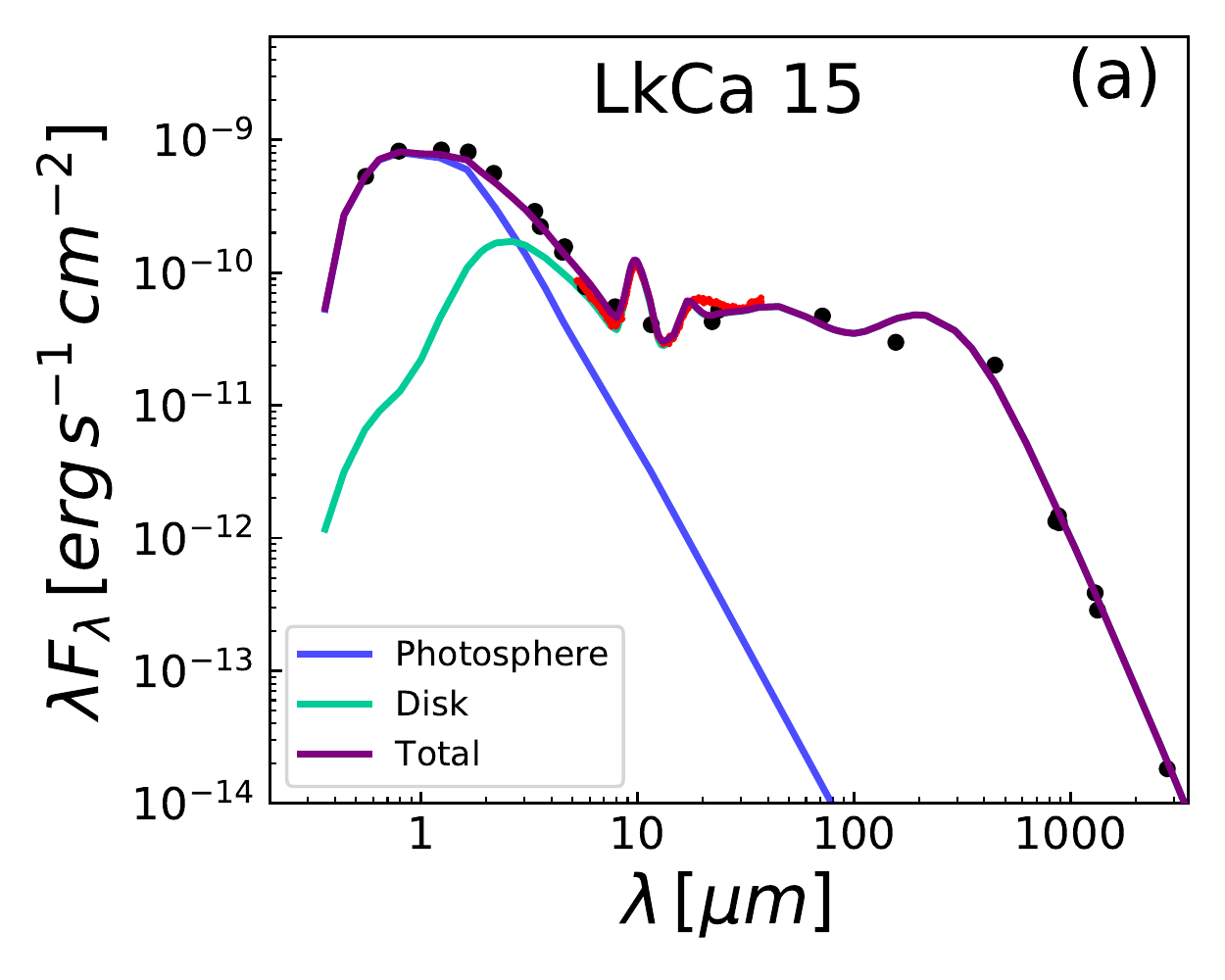}{0.32\textwidth}{}
          \fig{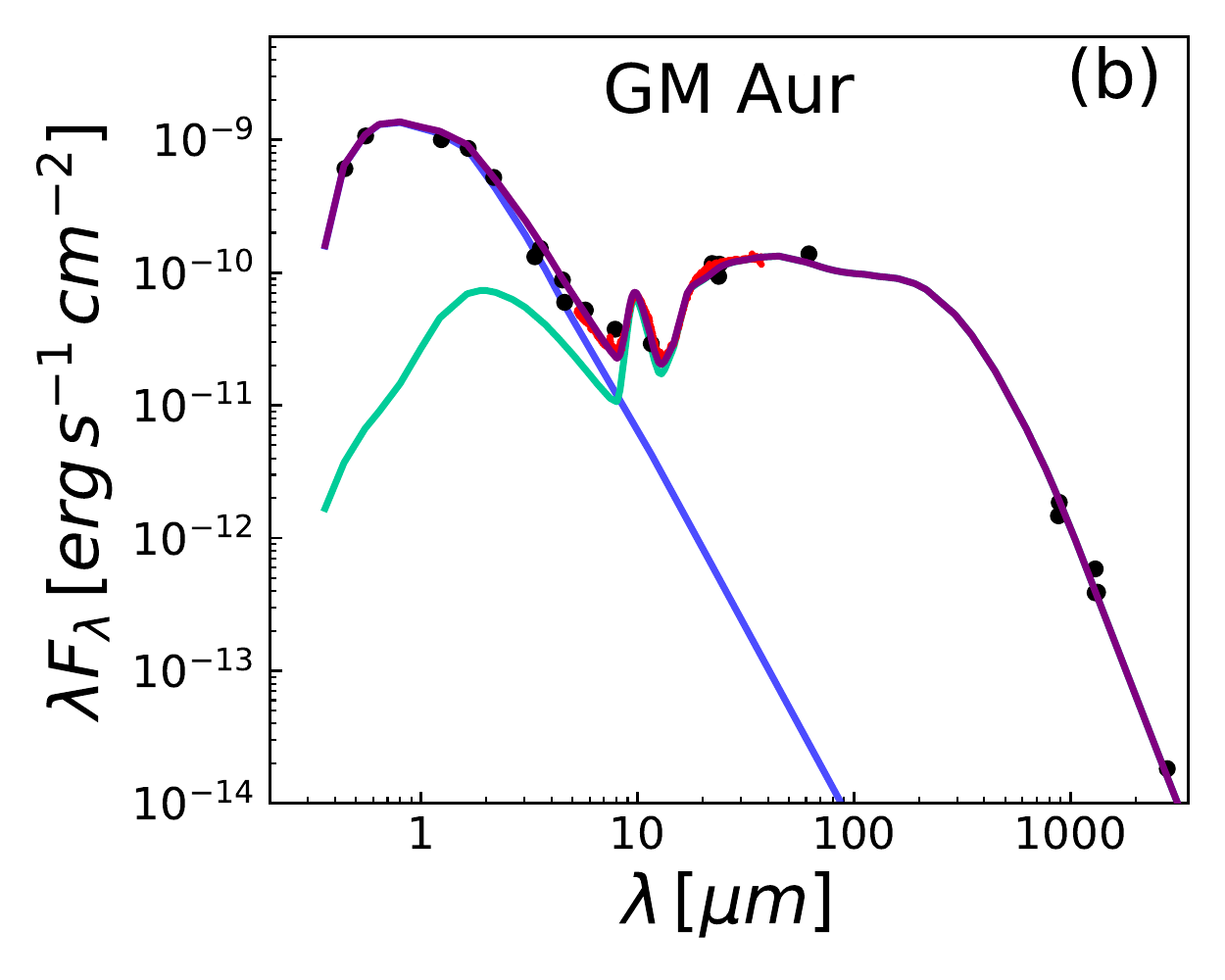}{0.32\textwidth}{}
          \fig{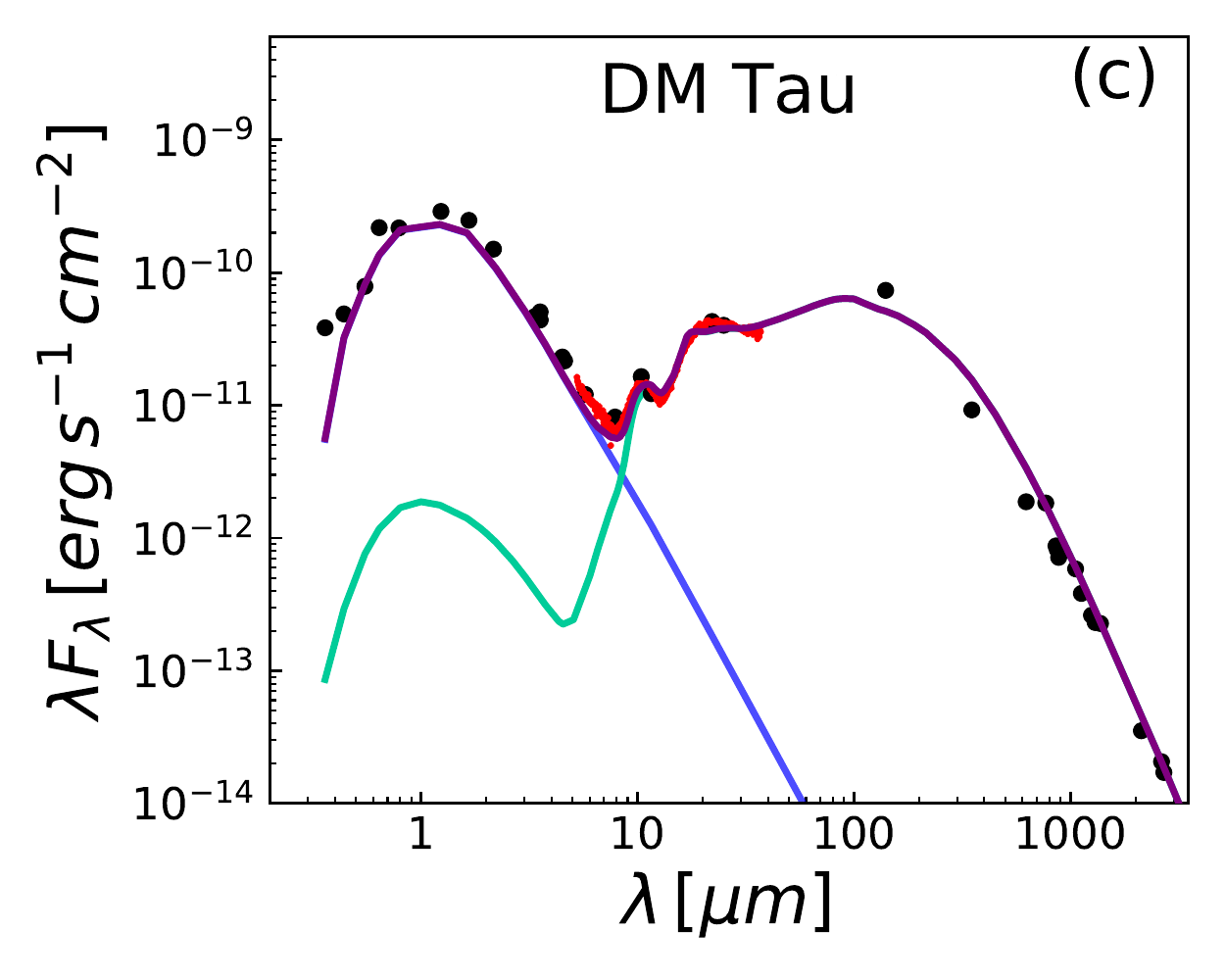}{0.32\textwidth}{}
          }
\gridline{\fig{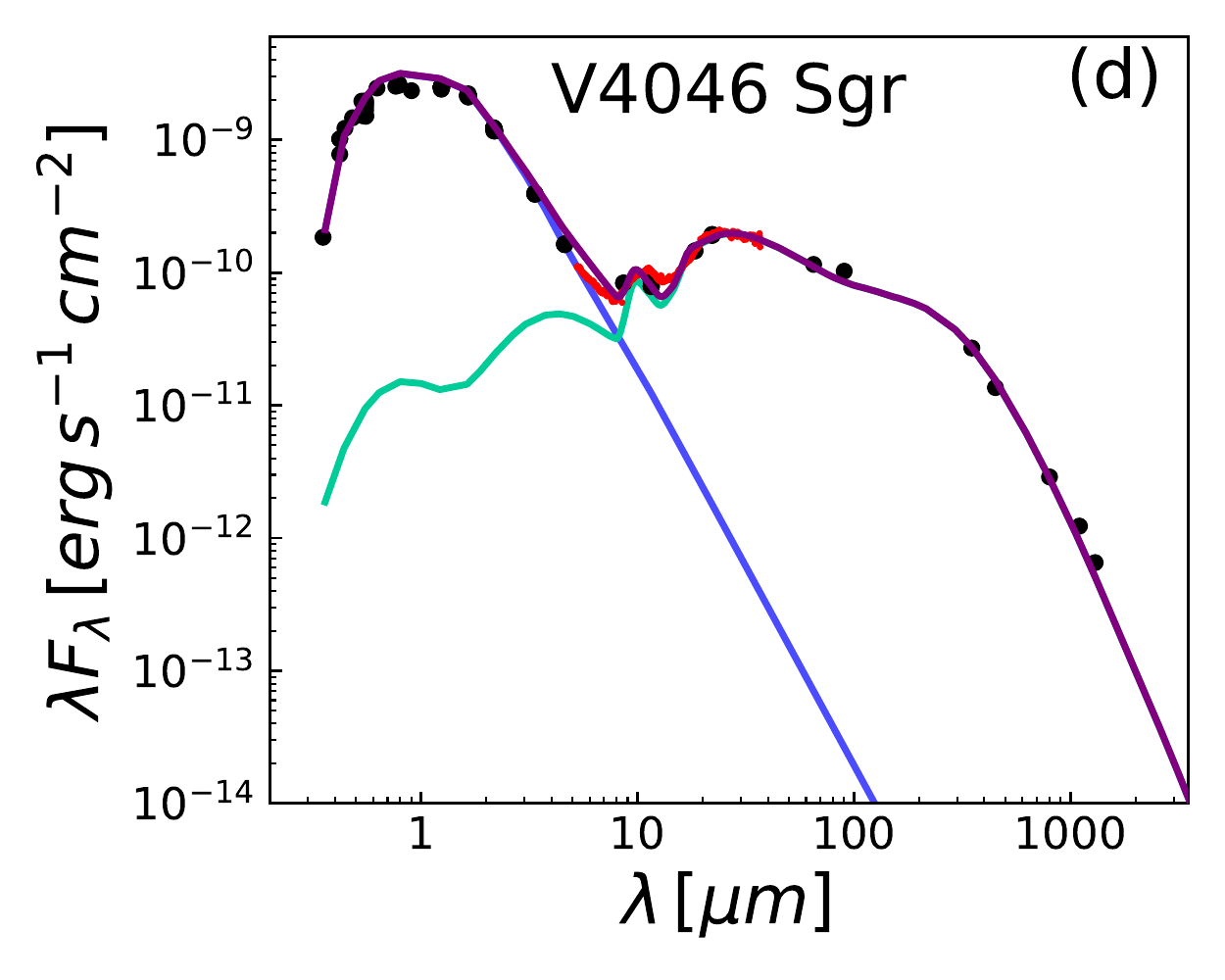}{0.32\textwidth}{}
          \fig{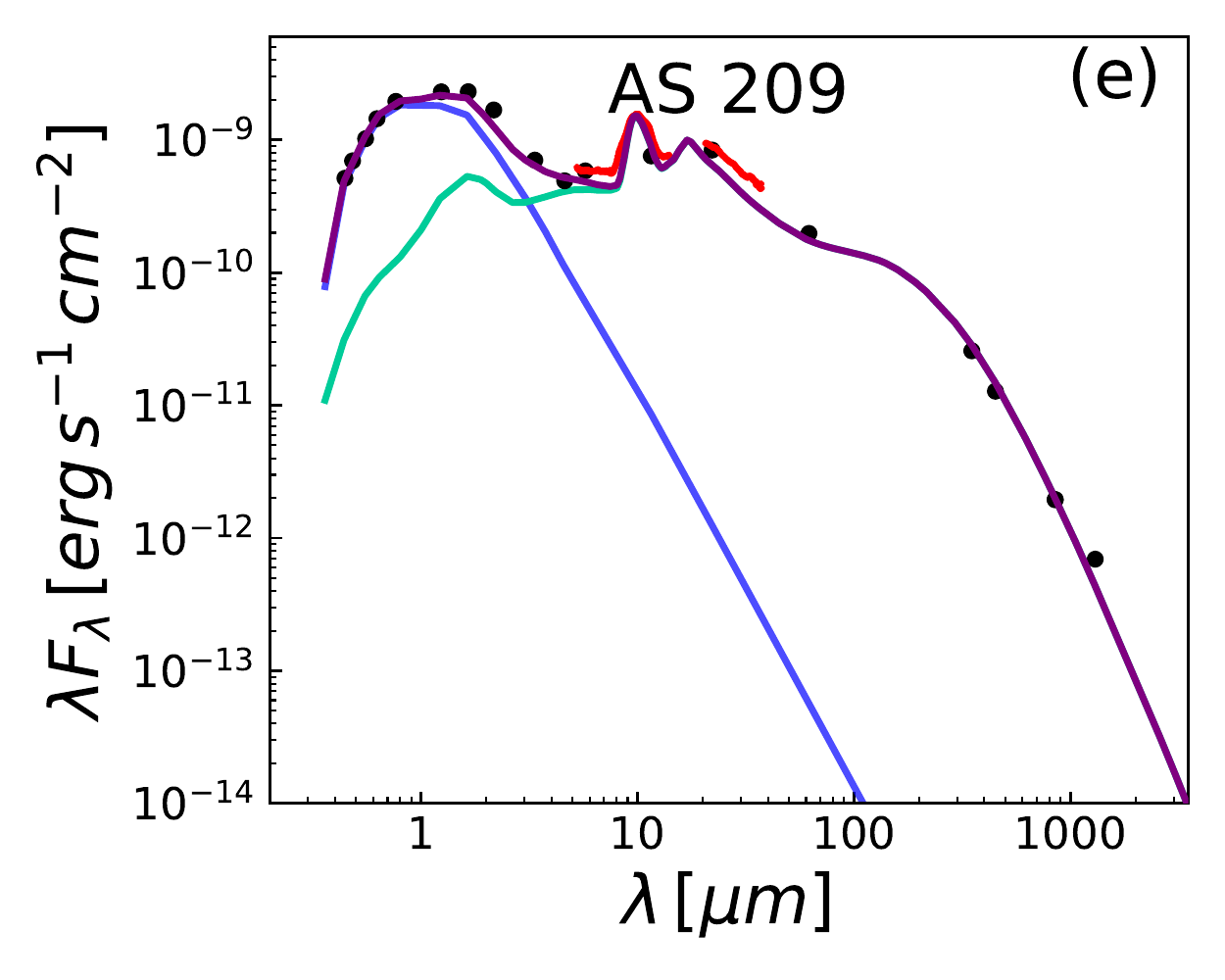}{0.32\textwidth}{}
          \fig{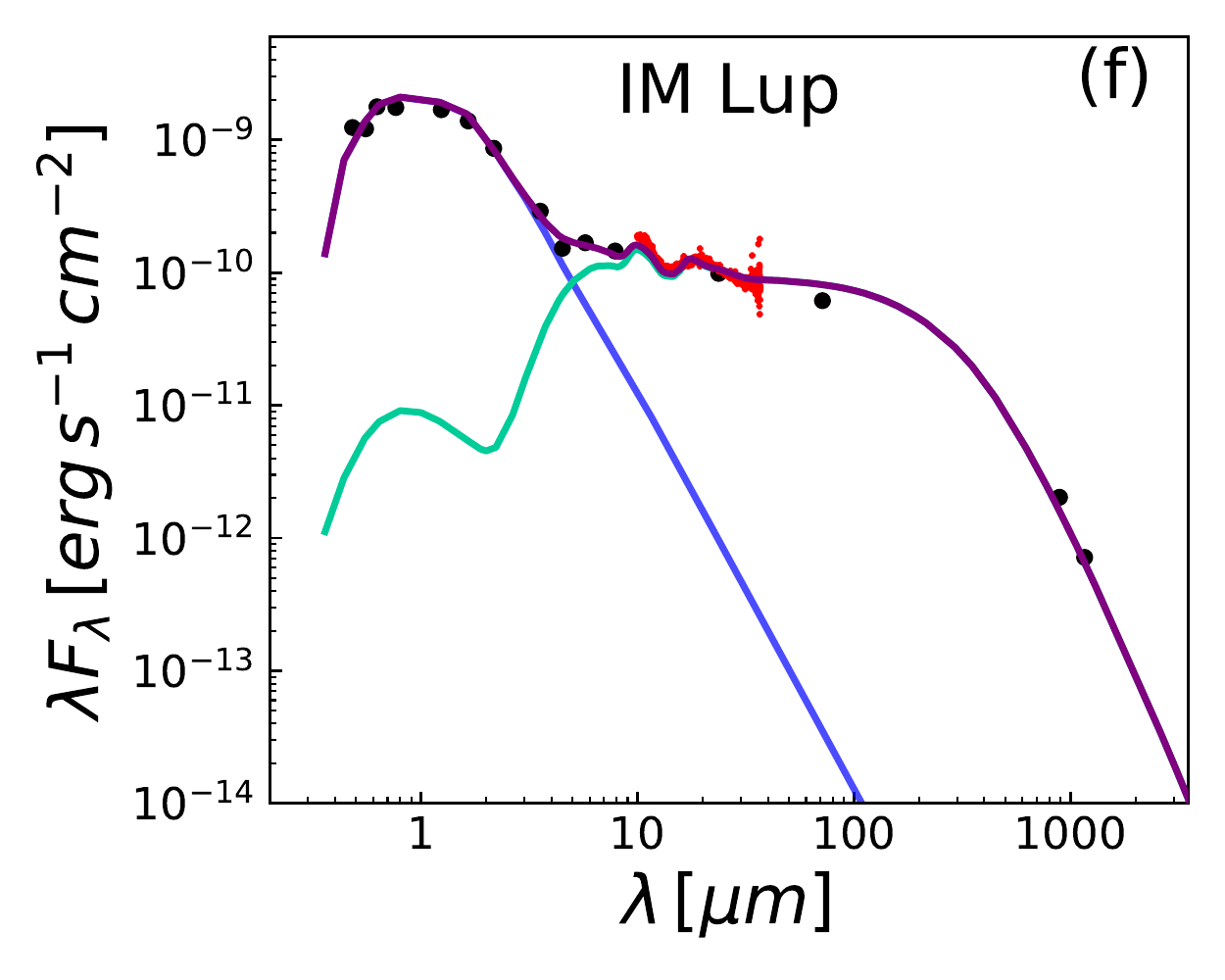}{0.32\textwidth}{}
          }
\caption{Spectral energy distributions (SEDs) of the six disks and the best fit models used to constrain the vertical disk temperature structures. The black points and error bars represent 
the measured photometry and the red line shows the IRS spectra. The blue lines represent the stellar
photosphere, the green lines are the disk models and the purple lines show the total emission of the
model. \label{fig:seds}}
\end{figure}

\begin{deluxetable*}{ccccccc|cccc|cccc}
\tabletypesize{\scriptsize}
\tablecaption{SED fitting parameters \label{tab:seds}}
\tablewidth{0pt}
\tablehead{
\colhead{} & \multicolumn{6}{c}{{\bf Outer Disk}} & \multicolumn{4}{c}{{\bf Inner Wall}} & \multicolumn{4}{c}{{\bf Optically Thin Dust Region}}\\
\colhead{Source} & \colhead{$\epsilon$} & 
\colhead{$\alpha$} & \colhead{T$_{wall}$} & \colhead{R$_{wall}$$^{a}$} & \colhead{H$_{wall}$} & \colhead{a$_{max}$} &
\colhead{T$_{wall}$$^{b}$} &\colhead{R$_{wall}$$^{a}$} &  \colhead{H$_{wall}$} & \colhead{a$_{max}$} &
\colhead{$\tau$} & \colhead{R$_{in}$} & \colhead{R$_{out}$} & \colhead{a$_{max}$} \\
\colhead{} & \colhead{} & \colhead{} & \colhead{(K)} & \colhead{(au)} & \colhead{(au)} &\colhead{($\mu$m)} & \colhead{(K)} & \colhead{(au)} & \colhead{(au)} & \colhead{($\mu$m)} & \colhead{} & \colhead{(au)} & \colhead{(au)} & \colhead{($\mu$m)} 
}
\startdata
LkCa 15 & 1.0 & 0.0001 & 95.0 & 68.75 & 1.8 & 0.25 & 1400 & 0.14 & 0.01 &1.0 & 0.02 & 0.3 & 5.0 & 0.25 \\
GM Aur& 1.0 &0.00045 & 120.0 & 29.95 & 1.9 & 3.0 & - & - & - & - & 0.007 & 0.1 &3.0 & 0.1\\
DM Tau & 0.5 & 0.0015 & 200.0 & 3.71 & 0.3 & 2.0 & - & - & - & - & - & - & - & -\\
V4046 Sgr &0.01 &0.0005 & 140.0 &12.04 & 0.7 &5.0 & - & - & - & - & 0.01 &0.2 &1.0 &4.0\\
AS 209 &0.001 &0.015 & 400.0 & 3.58 & 1.1 &0.25 &1400 &0.26 &0.1 &0.25 & - & - & - & -\\
IM Lup &0.001 &0.003 &500.0 &1.56 & 0.1 & 3.0 & - & - & - & - & - & - & - & -
\enddata
\tablecomments{\textsuperscript{a}R$_{wall}$ is calculated using T$_{wall}$ following \citet{dAlessio2005}.  \textsuperscript{b}We set T$_{wall}$ to an adopted dust sublimation temperature of 1400~K.}

\end{deluxetable*}

We tune 6-12 parameters for each object to fit its spectral energy distribution (SED; Table 5). 
The first 6 parameters in Table 5 pertain to the outer disk and are listed for each object. 
To achieve the best-fit, we test dust settling coefficient values  ($\epsilon$) of 1.0, 0.5, 0.1, .01, and 0.001 and vary accretion as parameterized by $\alpha$ between 0.00001 -- .02.  The inner disk edge or ``wall'', is modeled with temperature (T$_{wall}$) and the height (H$_{wall}$), which are varied to fit the SED.  The radius of the wall is calculated using T$_{wall}$ following \citet{dAlessio2005}. 
The maximum grain size in the disk atmosphere, a$_{max}$, was modeled as 0.1, 0.25, 1.0, 2.0, 3.0, 4.0, and 5.0 {\micron}.

In the cases of LkCa~15 and AS~209, the SED can be better characterized with an additional inner disk component with an optically thick wall located at the dust sublimation radius.  Here we adopt a dust sublimation temperature of 1400~K.
In addition LkCa~15, GM~Aur, and V4046~Sgr require some optically thin dust within their disk cavities to reproduce the SED. Following \citet{Espaillat2011}, we vary $\tau$, the vertical optical depth evaluated at 10~{\micron}, the inner and outer radii of the optically thin dust region (R$_{in}$, R$_{out}$), and a$_{\rm max}$.
The best-fit values are listed in
Table~\ref{tab:seds} and Figure.~\ref{fig:seds} shows that the best-fit models yield
excellent fits to the SEDs of the six disks.  The stellar photosphere, and optically thin dust reproduce well the optical and near/mid-IR data, while the tuned disk models provide good fits to the longer wavelength SED points. 

We note that this kind of SED fitting, especially to {\it Spitzer} IRS data points and the FIR wavelength region traced by {\it Herschel}, is highly sensitive to the vertical disk structure, the main goal of this study.
The main regulating parameter for the disk vertical structure in the DIAD models is  $\epsilon$, which describes the depletion of dust in the disk upper layers. All of our DIAD models are calculated using a mixture of two grain populations: small grains with a disk-specific max radius $a_{\rm max}$ (Table 5), and large grains with $a_{\rm max} = 1$\,mm. The latter grains are concentrated close to the disk midplane, within 10\% of the local gas scale height $H$, i.e. the height of the transition between the small and big grains, z$_{big}$=0.1$H$. We keep the vertically integrated dust-to-gas ratio constant, which implies that any missing dust in the disk upper layers as parameterized by $\epsilon$ has been moved, or settled to the midplane. Formally we define $\epsilon = \zeta_{\rm small}/\zeta_{\rm std}$, where $\zeta_{\rm small}$ 
is the dust-to-gas mass ratio 
in the upper layers and $\zeta_{\rm std}$ is the standard dust-to-gas mass ratio in the interstellar medium (see the detailed dust settling prescription in \citet{dAlessio2006}), i.e. the degree of settling increases as $\epsilon$ decreases. 

\begin{figure}
\includegraphics[width=\textwidth]{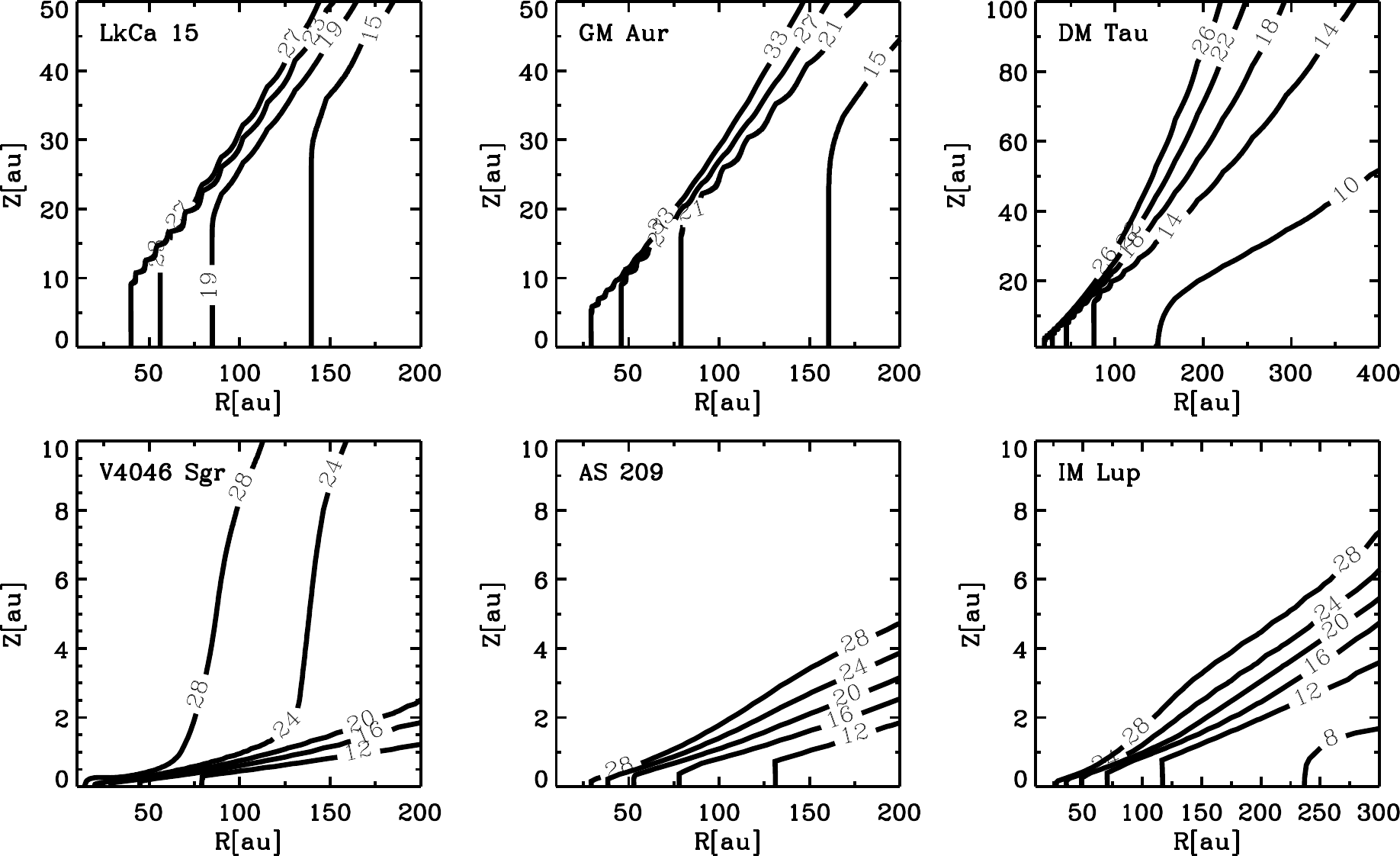}
\caption{Extracted disk temperature structures based on SED modeling. Note the different vertical scales in the upper vs lower panels. The disks in the upper panels all display substantial VIRaM layers, of $\sim$10~au around the CO and N$_2$ midplane the snowlines. By contrast the VIRaM layers in the disks in the lower panels are $<$1~au thick. \label{fig:vT}}
\end{figure}

Table~\ref{tab:seds} shows that in our sample the key fitting parameter $\epsilon$ is either 
$\epsilon \geq 0.5$ (LkCa~15, GM~Aur, and DM~Tau) or  
$\epsilon \leq 0.01$ (V4046~Sgr, AS~209, and IM~Lup). Figure.~\ref{fig:vT} shows the resulting vertical temperature structure of
the disks, and its profound dependence on
dust settling; the thickness of the VIRaM layer is more than an order of magnitude thicker in the disks with no settling, compared to the highly settled disks.
This dependence can be understood when considering the radiative transfer of stellar radiation through the disk. As demonstrated in \citet{dAlessio2006}, close to the disk midplane the disk temperature decreases with height since less stellar energy reaches the lower layers. At the disk height where the disk becomes optically thin to its own radiation 
the midplane becomes nearly isothermal. For disks with a high dust content in their upper layers ($\epsilon \geq 0.5$), this transition takes place at high disk altitudes and the VIRaM layer is therefore thick, extending to  
$z/r\approx0.2$. The result is nearly 
vertical temperature contours and therefore vertical snow surfaces for CO and N$_2$ from the disk midplane and up through most of the disk. For highly settled disks, stellar radiation can penetrate much deeper into the disk, and the transition to the optically thin, the isothermal disk layer takes place close to the disk midplane, 
at $z/r\approx 0.01$. As a result isotherms are highly inclined with respect to the surface normal through most of the disk, resulting in snow-surfaces that are highly inclined as well. 

We note that the thickness of the VIRaM layer is 
sensitive to the distribution of the small grains in the disk.
In the DIAD models, varying the 
parameter z$_{big}$, which is typically fixed to $0.1H$,
should also affects the disk vertical temperature structure,
as demonstrated in \citet{Qi2011}.
 Besides gravitational settling, dust grain dynamics (e.g., 
radial drift, dust fragmentation), which are not yet considered
in the DIAD models, should also change the spatial distribution
of the small grains.  We do not explore 
these effects here, as we do not expect a more detailed 
treatment will result in qualitative changes
to the VIRaM structure of these disks.

\subsection{Snowline(s) characteristics}

\nnhp{} emission morphologies in our disk sample all present as a ring structure, and the radial location of the inner edge of the ring should provide information about the CO snowline location. An initial estimate of the inner ring edge can be obtained through the detection of the \nnhp emission at the highest velocity channels which corresponds to the emission from the innermost radii considering the disk in Keplerian rotation. By measuring the velocity difference $\delta V$ (in unit of km s$^{-1}$) between the systemic velocity and the velocity in the most blue-shifted channel \footnote{Blue-shifted emission (reference to 279.5117880 GHz) is less affected by the hyper-fine components of \nnhp{} $3-2$.} with $>3\sigma$ detection (See Figures~\ref{fig:lkca15channel}-\ref{fig:imlupchannel} in the Appendix A), 
we derive the inner edge radii R (in unit of au) as $100 \times M_{star} (\frac{2.98 \sin{i}}{\delta V})^2$ where M$_{star}$ is the stellar mass in unit of M$_{\odot}$ and $i$ is the inclination of the disk as listed in Tables~\ref{tab:stellar} and ~\ref{tab:cont}. Inserting the values, we obtain the initial estimates of the inner edge of \nnhp{} ring as 51, 41, and 72 au in the disks of LkCa~15, GM~Aur, and DM~Tau (first group) and 33, 31, and 59 au in the disks of V4046~Sgr, AS~209, and IM~Lup (second group). As shown in Figure.~\ref{fig:n2hpsnow}, these values can be treated as the initial estimates of the CO snowline locations for the first group and the upper limits for the second one.  Due to the lack of a thick VIRaM layer in the disks of the second group, the N$_2$ snowline cannot be constrained from the \nnhp{} emission profiles. 
We note that the uncertainty in the velocity offset analysis described
here is hard to determine, especially for the low inclination
disk systems as it suffers more confusion from the hyper-fine
components of the N$_2$H$^+$ line.
More rigorous constraints on the CO and N$_2$ snowline locations in the disk of the first group can be obtained using  $\chi^2$ analysis of the visibilities in the $(u,v)$-plane. 

\begin{figure}
\includegraphics[width=\textwidth]{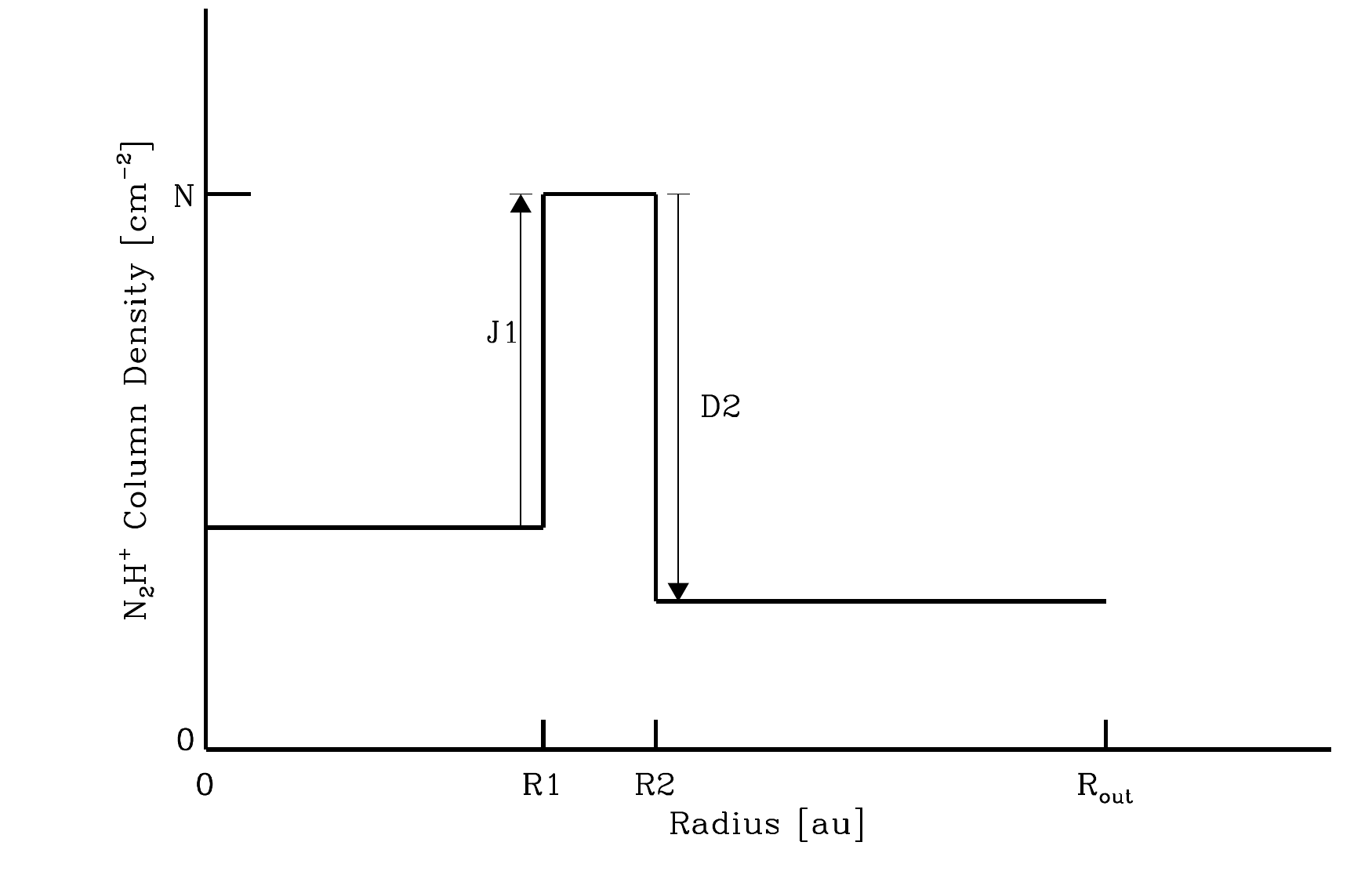}
\caption{Schematic of the column density ``jump and drop" model used to derived CO and N$_2$ snowline locations for disks with a thick VIRaM layer. R1 and R2 correspond to the CO and N$_2$ midplane snow-line locations, respectively, J1 is the N$_2$H$^+$ enhancement following CO freeze-out in the midplane, and D2 the N$_2$H$^+$ drop exterior to the N$_2$ snowline, where N$_2$H$^+$ only exist in a thin, elevated disk layer. \label{fig:modelling}}
\end{figure}

It is a remarkable confirmation of disk chemistry theory that the disks in \S3.3 that are modeled to have little settling and therefore a thick VIRaM layer perfectly overlap with the disks that present narrow and well-defined N$_2$H$^+$ rings. In these three disks
we expect near vertical CO and N$_2$ snow surfaces extending from the midplane 
and therefore the inner and outer edges of the \nnhp{} ring should well isolate the CO and N$_2$ midplane snow lines,
as shown schematically in Figure.~\ref{fig:n2hpsnow}. To derive snowline locations from the \nnhp{} emission we use a parametric abundance model, with
a ``jump and drop" radial 
column density profile of \nnhp{} (Figure.~\ref{fig:modelling}) 
to simulate the effects of freeze-out of gas-phase CO (producing the jump) and freeze-out of N$_2$ (producing the drop).
The model parameters are the peak 
column density of \nnhp{}, N$_p$ 
and the column density `jump' factor J1 and `drop' factor D2 at the corresponding inner and outer edges of the bright ring, R1 and R2, as well as an outer radius $R_{\rm out}$.

In the vertical dimension, we follow the methodology of \citet{Qi2013c, Qi2015b} 
and assume that the \nnhp{} abundance is  constant between the disk surface ($\sigma_s$) and midplane ($\sigma_m$) boundaries at each radius. These boundaries are described in terms of $\Sigma_{21}=\Sigma_H /
(1.59\times10^{21} {\rm cm}^{-2})$, where $\Sigma_H$ is the hydrogen column
density (measured downward from the disk surface) in the adopted physical
model. We fix the boundary values to be 3.2 and 100, appropriate for a disk 
midplane tracer like \nnhp{} according to 
e.g., the chemical models of \citet[Figure. 8]{Aikawa2006}.  
The abundance at any one radius is then completely described by the \nnhp{} column density model described above and the given disk density model.

We use the 2D Monte Carlo software RATRAN \citep{Hogerheijde2000} to calculate the radiative transfer and molecular excitation.
\nnhp{} has a hyper-fine structure due to the nuclear quadruple moment of
$^{14}$N 
and relative populations between the hyper-fine levels are
assumed to be in LTE. The collisional rates are adopted from \citet{Flower1999} based on HCO$^+$ collisional rates with H$_2$, which are taken to be the same as for \nnhp{}. The molecular data files are retrieved from the Leiden Atomic and
Molecular Database \citep{Schoier2005}.

Because the radiative transfer calculation is very time consuming, we separate the fitting of the \nnhp{} distribution and abundance parameters (N$_p$, J1, D2, R1, R2, and $R_{\rm out}$) and the disk geometric and kinematic parameters (disk inclination $i$, disk position angle P.A., and the stellar mass). We fix $i$ and disk P.A. by using the best-fit values from the continuum analysis (Table~\ref{tab:cont}). We adopt the stellar masses 1.0, 1.3, 0.5 M$_{\odot}$ for LkCa~15, GM~Aur, and DM~Tau, respectively, from the literature \citep{Andrews2018a} for initial calculations. 
For each disk, we fit for the \nnhp{} distribution and abundance parameters by running a large grid of parametric models \footnote{Specifically, R1 and R2 are fit with a grid with interval of 5 au.}, calculating the predicted \nnhp{} emission,
 and then comparing simulated visibilities with observed ones.
The best-fit parameter estimates are obtained by minimizing
$\chi^2$, the weighted residual of the complex visibilities
measured at the ($u,v$)-spacings sampled by ALMA. 
To obtain the best fitting results, we fit the stellar masses
for the 3 sources after obtaining the initial best-fit distribution parameters.
Using the newly derived stellar masses, we repeat the fitting of the \nnhp{} distribution and abundance parameters. Finally the stellar masses are refit and confirmed with no more changes and the values are listed in 
Table~\ref{tab:stellar}. 
The final $\chi^2$ values and the best-fit image quality are much improved.

\begin{deluxetable*}{lcccccccc}
\tablecaption{\nnhp{} fitting results from drop and-jump model\label{tab:fit}}
\tablewidth{0pt}
\tablehead{
\colhead{Source} & \colhead{R1} & \colhead{T1$^a$} 
& \colhead{J1} & \colhead{R2} & \colhead{T2$^a$} & \colhead{D2} &
\colhead{R$_{out}$} & \colhead{N$_{p}$} \\
\colhead{} & \colhead{(au)} & \colhead{(K)} & \colhead{} & \colhead{(au)}
& \colhead{(K)} & \colhead{} & \colhead{(au)} &\colhead{(10$^{12}$ cm$^{-2}$})}
\startdata
LkCa 15 & 58$^{+6}_{-10}$ & 21--25 & 10$^{+15}_{-4}$ & 88$^{+6}_{-4}$ & 18--19 & 6.3$^{+0.3}_{-0.7}$ & 360$\pm$20 & 5.3$\pm$0.2 \\
GM Aur & 48$^{+10}_{-8}$ & 24--28 & 6.3$^{+25}_{-2}$ & 78$^{+4}_{-6}$ & 20--22 & 4.0$^{+1.0}_{-0.5}$ & 320$\pm$20 & 3.1$\pm$0.2 \\
DM Tau & 75$^{+10}_{-30}$ & 13--18 & 3.2$^{+1.8}_{-1.0}$ &  145$^{+15}_{-10}$ &  9--10 &4.0$\pm$0.5 & 420$\pm$20 & 1.3$\pm$0.1 \\
\enddata
\tablecomments{\textsuperscript{a} The disk midplane temperatures (presented as ranges) correspond to the locations of R1 and R2, i.e. the snowline temperatures for CO and N$_2$ in the disks.}
\end{deluxetable*}

\begin{figure}
\includegraphics[width=\textwidth]{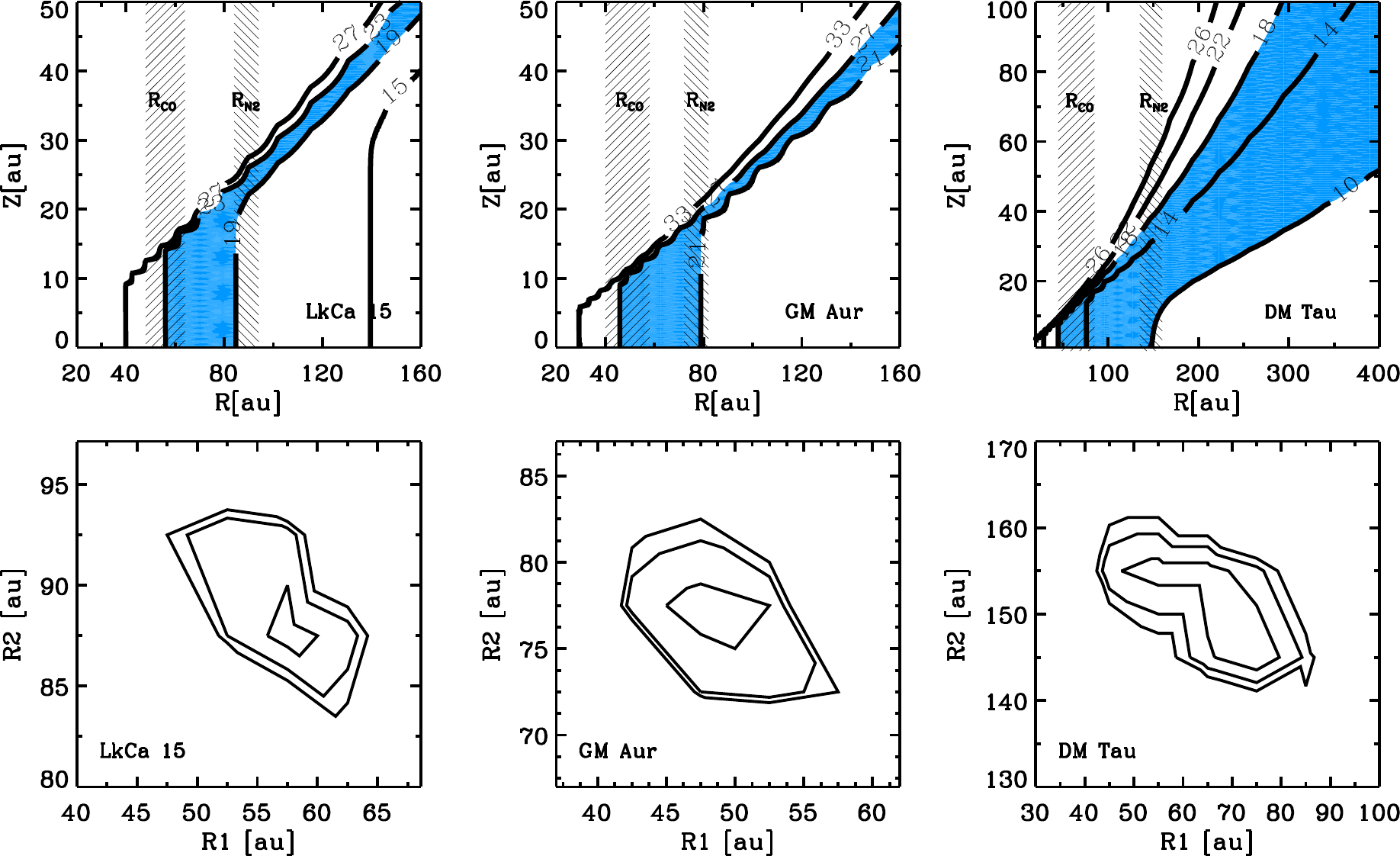}
\caption{{\it Top panels:} CO and N$_2$ snowlines and and snow-surfaces shown on top of disk temperature profiles. The striped regions show the constraints on the inner and outer N$_2$H$^+$ edges from the jump-and-drop model, interpreted as the CO and N$_2$ midplane snowline locations. The blue filled regions mark the temperatures between CO and N$_2$ freeze-out, extracted from a second grid of parametric models. Note the excellent agreement. {\it Lower panels:} $\chi^2$ surfaces showing model constrains on R1 (inner edge) and R2 (outer edge) in jump-and-drop model for each disk. Contours correspond to the 1--3 $\sigma$ uncertainties. \label{fig:modelfit}}
\end{figure}

\begin{figure}
\includegraphics[width=\textwidth]{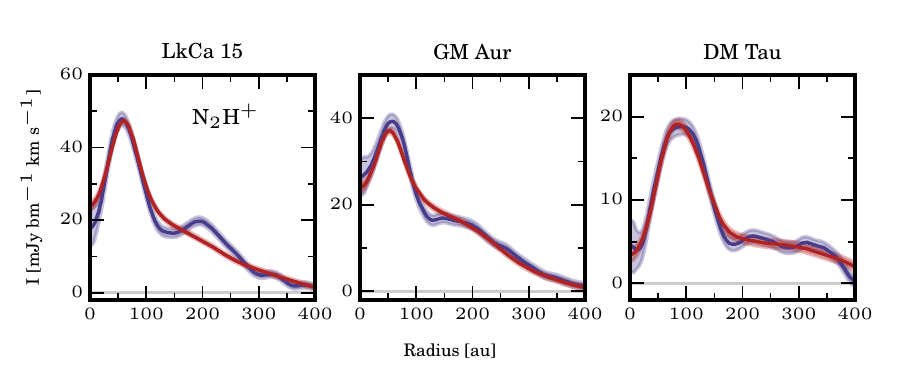}
\caption{Observed \nnhp{} $3-2$ profiles vs simulated observations of best-fit models in the jump-and-drop model framework illustrated in Figure. 6. \label{fig:modelprofiles}}
\end{figure}

Table ~\ref{tab:fit} shows the best-fit parameters for R1, R2,J1,
D2 and N$_p$. Figure.~\ref{fig:modelfit} shows the $\chi^2$ plot and 
the snowline locations overlapped on the temperature contours of the
disks. We find CO snowlines at 48--75~AU and N$_2$ snowlines at 78--145 AU in the three disks. The derived CO snowlines are consistent with the initial estimates. Figure~\ref{fig:modelprofiles} shows the deprojected profile of 
the observed \nnhp{} emission compared with the best-fit models. 
The best-fit models match the observations very well. 

We note that the uncertainties on R1 and R2 of these disks, derived from the $\chi^2$ contours, should reflect
the sharpness of the ring edges. The uncertainties are typically $\sim10-40\%$ for R1 and  $\sim5-10\%$ for R2 with respect to the best fit value. Table~\ref{tab:fit} indicates that 
the edges on the disks of LkCa~15 and GM~Aur are sharper than those of DM~Tau. 
This could relate to a smaller $\epsilon$ for dust settling for DM~Tau compared to the other two disks. 
Observations on a larger sample of disks are needed to confirm whether the sharpness of the edges is correlated with $\epsilon$.
Turbulent diffusion in disks might also play an 
important role on the sharpness of the emission edges \citep{Owen2014},
which needs to be further explored. 

Based on comparison between R1 and R2, and the disk temperature structures we can derive snowline temperatures for CO and N$_2$ in the three disks as shown in Table~\ref{tab:fit}. We find CO snowline temperatures of 21--25, 24--28 and 13--18~K, and N$_2$ snowline temperatures of 18--19, 20--22, and 9--10~K for LkCa~15, GM~Aur and DM~Tau, respectively. The low values obtained for DM~Tau are noteworthy and are discussed further below. 

Perhaps simplistically we expect that the same freeze-out and desorption equilibrium that is setting the snowline temperature in the midplane should also set it along the snow surfaces throughout the disk. A complementary model approach to the above `Jump-and-drop' model is then to assume a constant \nnhp{} abundance between two temperature boundaries corresponding to CO and N$_2$ freeze-out.
To evaluate the robustness of the results obtained from the jump-and-drop model, we therefore ran a second grid of parametric models characterized
by the CO freeze-out temperature T$_{\rm{CO}}$, the N$_2$ 
freeze-out temperature T$_{\rm{N_2}}$, and 
the fractional abundance of \nnhp{}. 
The best-fit T$_{\rm{CO}}$ and T$_{\rm{N_2}}$\footnote{There is no attempt to determine uncertainties on T$_{\rm{CO}}$ and T$_{\rm{N_2}}$ for the complementary model.} are 23 and 19 K for LkCa~15, 
27 and 21 K for GM~Aur, and  18 and 10 K for DM~Tau, in perfect agreement with the results obtained from the jump-and-drop models. The locations of these temperature regions are shown by the blue shades 
in Figure.~\ref{fig:modelfit}, demonstrating excellent agreement between this set of models and the jump-and-drop models used to establish snowline locations. 

\section{Discussion} \label{sec:disc}

\subsection{{\rm CO} and {\rm N$_2$} freeze-out temperatures}

There are two previously derived CO snowline temperatures from \nnhp emission modeling: 17~K in the disk of TW Hya, and 25~K in the disk of HD 163295 \citep{Qi2013c,Qi2015b}. The low value for TW Hya has been contested, however, and an analysis using CO isotopologues instead resulted in a CO snowline at 27~K \citep{Zhang2017}. In the disk warm molecular layer, the freeze-out termperature of CO is constrainted to be around 21 K toward the TW~Hya \citep{Schwarz2016} and IM~Lup \citep{Pinte2018} disks. 
Our new CO snowline temperatures of  13--18~K (DM~Tau), 21--25~K (LkCa~15), and 24--28~K (GM~Aur) add to this list and show that there is a real range of CO snowline temperatures, also among T Tauri disks.

The ranges of observed CO and N$_2$ snowline temperatures for GM~Aur and LkCa~15 compare well with expectations from laboratory experiments \citep{Collings2004,Oberg2005,Bisschop2006,Fayolle2016}. Measured CO binding energies range from 870~K (pure CO ice) to 1300~K (adsorbed onto compact H$_2$O ice) corresponding to snowline temperatures of $\sim$21--32~K, respectively, assuming a midplane n$_{\rm H}$ density of 10$^{10}$ cm$^{-3}$ and a CO abundance of $5\times10^{-5}$ with respect to n$_{\rm H}$.  Measured N$_2$ binding energies range from 770~K to 1140 for analogous ices, corresponding to snowline temperatures of $\sim$18--27~K, respectively, assuming the same midplane n$_{\rm H}$ density as above, and a N$_2$ abundance of $3\times10^{-5}$ with respect to n$_{\rm H}$ (assuming that 95\% of the N atoms are in N$_2$). The extracted LkCa~15 and GM~Aur CO and N$_2$ snowline temperatures suggest that both volatiles reside on moderately water-rich ice grains, if the disks are effectively static. Pebbles are however subject to radial drift on timescales similar to sublimation, which can move snowlines inwards, to higher disk temperatures than expected \citep{Piso2015}. If drift is important in these disks, or if the disks are a few degrees warmer than modelled, CO and N$_2$ could be sublimating from close to pure ice layers. 

The DM~Tau CO and N$_2$ snowline temperatures are considerably lower than the lowest expected value; the contrast between expected and observed N$_2$ snowline temperatures is almost a factor of 2!  There are possible explanations for such low freeze-out temperatures, including radial mixing, 
turbulent, vertical mixing \citep{Aikawa2007}, and photodesorption \citep{Hersant2009}, but all are highly speculative considering the large opacities and low turbulence levels expected in disk midplanes \citep[e.g.][]{Teague2016,Flaherty2018}. Photodesorption has been previously invoked as an explanation for observations of  cold CO gas  in the GM~Aur, LkCa~15, and DM~Tau disks \citep{Dartois2003, Pietu2007}, and may play a larger role in setting the division of volatiles between gas and grains in the DM~Tau disk compared to the other two disks. Another possible explanation is that the disk is actually substantially warmer than the best-fit DIAD model. To distinguish between these different explanations we need direct measurements of the N$_2$ gas temperatures between the CO and N$_2$ snowlines. Resolved observations of a second \nnhp{} transition with a substantial difference in upper energy level from the existing $3-2$ transition, will be essential to provide the model-independent measurements of the temperature range at which CO and N$_2$ snowlines occur.

Finally it is illustrative to compare the ratios of observed and expected CO and N$_2$ snowline temperatures. In laboratory experiments the N$_2$ sublimation energy is consistently 10\% lower than the CO sublimation temperature, when considering identical ice environments. Once again the LkCa~15 and GM~Aur observations are consistent with expectations -- snowline temperature ratios range between 0.7 and 0.9 for both disks -- and DM~Tau is not. In the latter case the CO/N$_2$ snowline temperature ratio varies between 0.5 and 0.77. Such low ratios can be achieved in the laboratory if  CO sublimation from a water-rich ice is compared with N$_2$ sublimation from a water-poor ice. Given that the DM~Tau temperature profile is correct,  the DM~Tau results suggest that CO and N$_2$ sometimes reside in different ice environments in the same disk, where e.g. CO is in a more strongly bound ice, while N$_2$ is frozen out in a weakly bonding ice layer. There may thus not be a fixed ratio between the CO and N$_2$ snowline locations in disks. We note, however, that if the adopted temperature profile is off by only two degrees at the N$_2$ snowline we could not rule out that both CO and N$_2$ are present in similar, hypervolatile ice environments.

\subsection{Snowlines and double-ring dust substructures in disks}

Millimeter observations of protoplanetary disks at high angular resolution
have revealed a wealth of substructures
\citep[e.g.][]{ALMA2015, Andrews2012,Isella2016,Long2018,Andrews2018b}.
Many of these structures are concentric and axisymmetric, e.g. gaps and rings. 
The snowlines of major volatile species
may play a role in creating these features, through 
rapid particle growth by condensation \citep[e.g.][]{Ros2013, 
Zhang2015,Pinilla2017}, or aggregate sintering \citep{Okuzumi2016}, or pile-ups of material due to increased fragmentation \citep{Stammler2017}. If the latter mechanism dominates, the increase in surface density around snowlines will be seen as bright rings in millimeter observations. The more abundant the volatile species is, the effect is stronger. Therefore 
we should expect a double-ring system in the outer disk associated with the CO 
and N$_2$ snowlines.

The relationship between snowline locations and disk sub-structure has recently been tested in large samples with 10s of disks. In particular, \citet{Long2018} and \citet{Huang2018b} found no obvious correspondence between 
the locations of the substructures and the disk midplane temperatures, and inferred that major snow lines in mature disks do not play an important role in regulating observed sub-structures. 
These conclusions rely on two assumptions, however, that disk midplane temperature structures can be well approximated using simple models, and that the same snowlines generally occur at the same disk temperatures. In Section~\ref{sec:results} we showed that snowline temperatures can vary by up to a factor of two and this range may in reality be even larger, since we do not account for snowline locations in settled disks. It is difficult to know the temperature structure of a disk in detail, but the chemical structure (i.e. where the \nnhp{} emission lies) is perhaps a more robust way of isolating the CO and N$_2$ snowlines.

\begin{figure}
\includegraphics[scale=1]{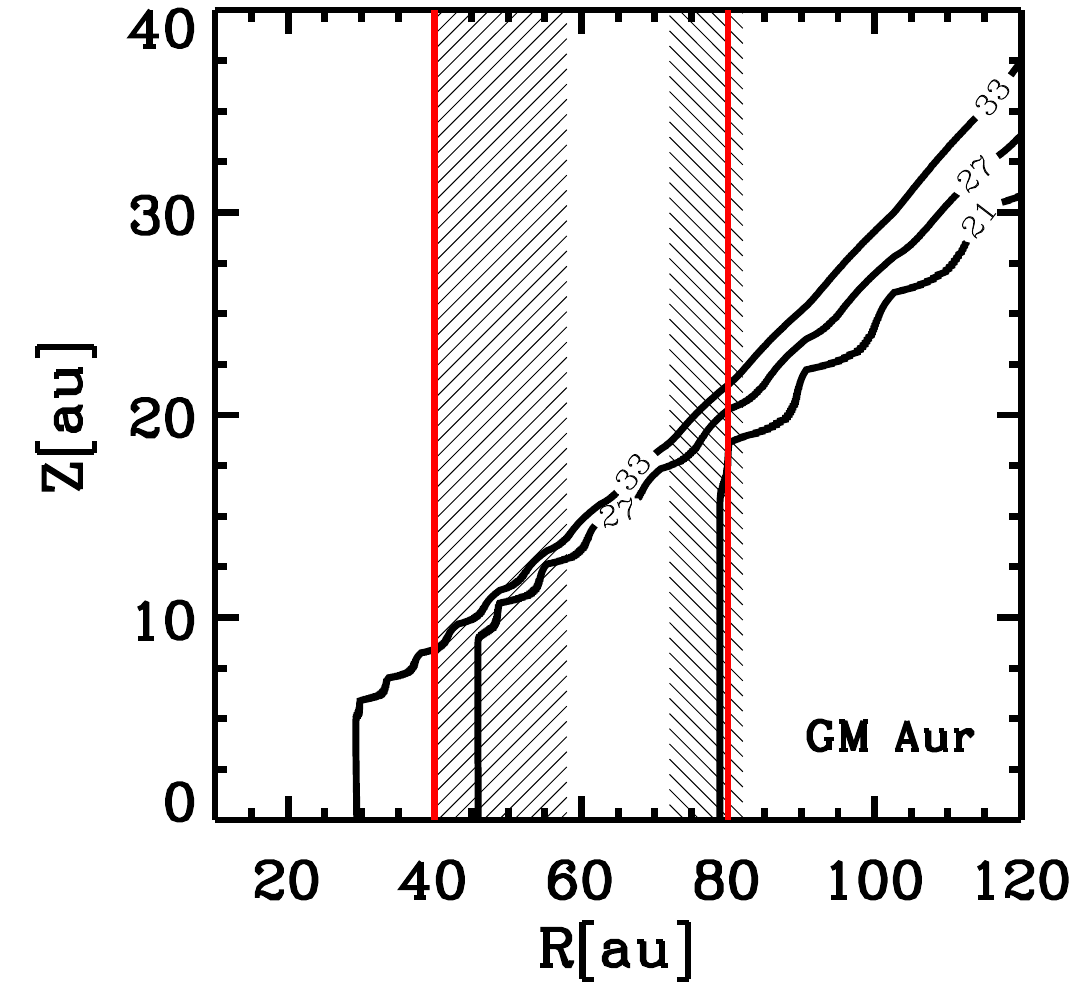}
\caption{The dust ring locations determined in the GM~Aur disk \citep{Macias2018} in red lines and the constraints on the CO and N$_2$ snowline locations in striped regions shown on top of its disk temperature profiles in contours. \label{fig:gmaurdust}}
\end{figure}

One of our disks, GM~Aur, is observed at high enough resolution to resolve its ringed sub-structure, and in this case we do not need to make assumptions about snowline locations, but can rather test directly whether there is a relationship between snowlines and dust rings.
High resolution ($\sim$0.2$''$) observations of
dust continuum emission from GM~Aur \citep{Macias2018} 
revealed two bright rings at 40 and 80 au (Figure.~\ref{fig:gmaurdust}), which can be compared to our extracted CO and N$_2$ snowline locations of 40--58 AU and 74--84~AU, respectively. 
This coincidence is suggestive, but needs to be tested for more disks. 
So far, high angular resolution ($\sim$40 mas observations of DM~Tau don't appear to show any double-ring dust structure associated  with the CO and N$_2$ snowline locations \citep{kudo2018}. However their integrations on source were too short to detect the little  ``bump" around 100 AU from the azimuthally averaged radial profile in our deprojected image of DM~Tau (Figure~\ref{fig:profiles}). 
So long-baseline continuum observations with deep integrations are needed for 
the DM~Tau and LkCa~15 disks to define better their detailed 
sub-structure.
The presence of similar double-ring
systems coincident with snowline locations in all disks would provide
evidence of a close relationship between snowline locations and dust sub-structure. 

Finally we note, that whether a disk has vertical snow surfaces, may affect how much midplane snowlines 
change the coagulation, fragmentation and sintering efficiencies of pebbles. All existing models assume
a vertically isothermal disk and hence vertical snow surfaces \citep[e.g.][]{Okuzumi2016, Stammler2017}, 
and flatter snow surfaces may reduce the effect of snowline crossings on grain growth and destruction. It is therefore possible that there are two populations of disks with sub-structures, one where some or all dust rings are caused by snowline-related processes, and a second, settled disk population where snowline locations do not affect the emergence of dust sub-structure.

\subsection{Disk temperature structure and uncertainty in gas CO measurements}
\begin{figure}[ht!]
\centering
\includegraphics[width=0.42\textwidth]{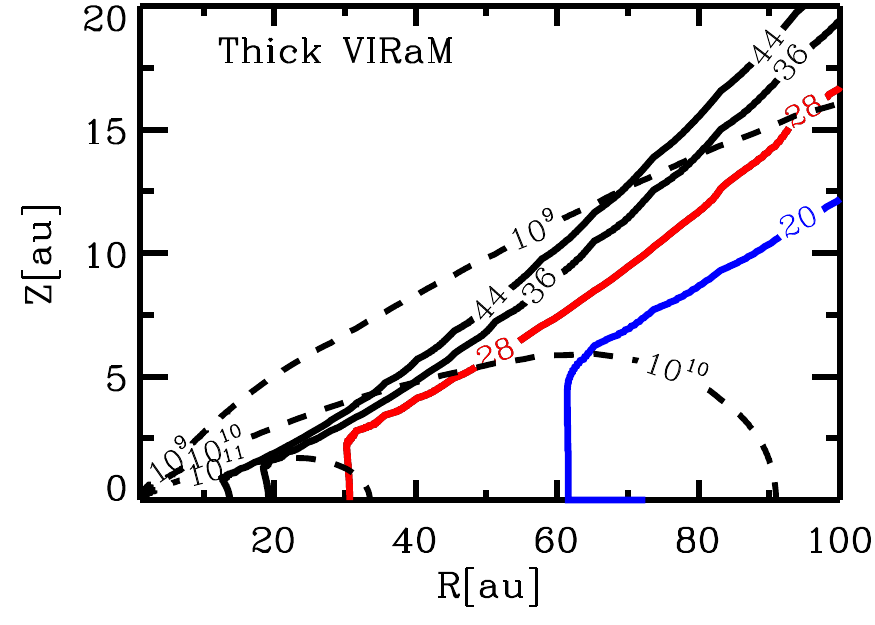}
\includegraphics[width=0.42\textwidth]{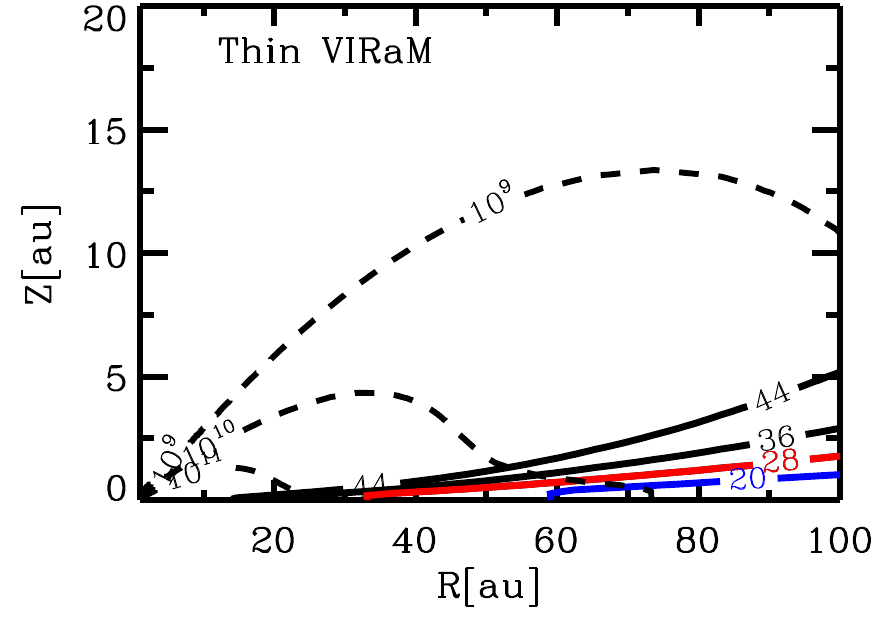} \\
\vskip 5mm
\includegraphics[width=0.42\textwidth]{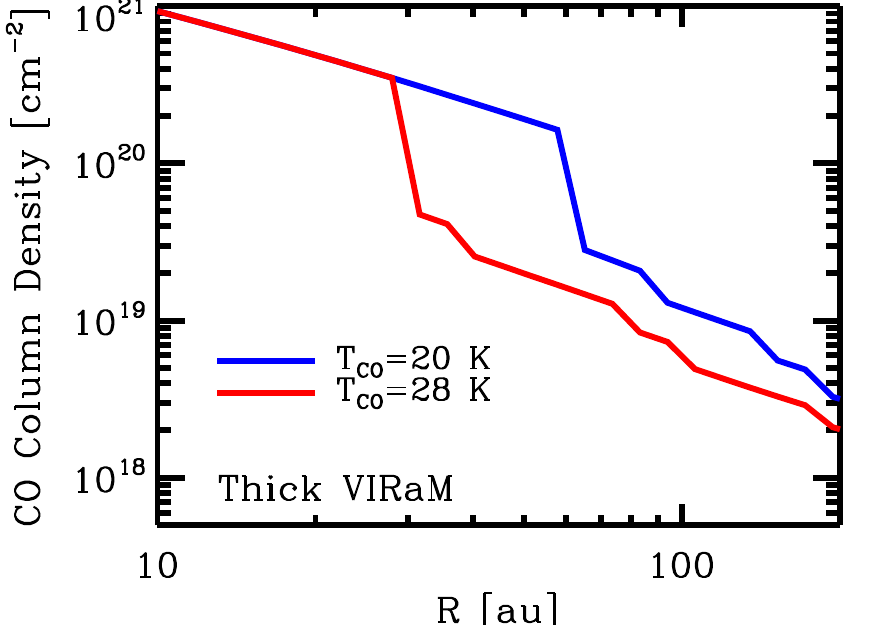}
\includegraphics[width=0.42\textwidth]{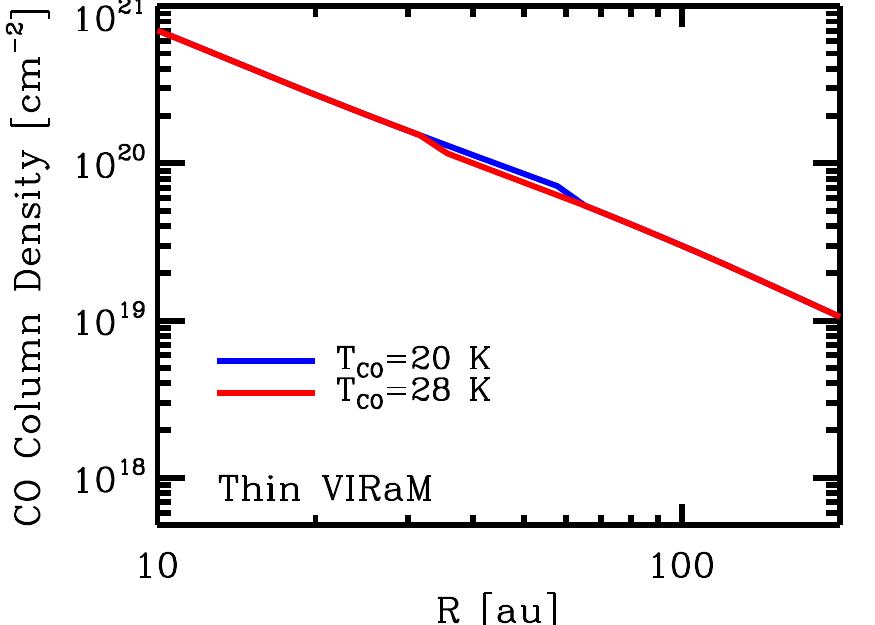} \\
\caption{\label{fig:comodels}{\it Top panels:} Temperature and density profiles of the models with a thick or thin VIRaM layer. {\it Bottom panels:} The predicted CO radial column density profiles for both models, assuming T$_{\rm CO}$=28 K (red line) and 20 K (blue line).}
\end{figure}

An additional implication of this work regards the total disk gas mass. The determination of protoplanetary disk masses is fundamental for understanding the formation and evolution of disks and planets. The observation of CO and its isotopologues has been used as a measure of
the total disk gas content \citep[e.g.][]{Williams2014}. However there are many sources of uncertainty
in gas mass measurements from CO observations, e.g. chemical sequestration \citep{Bergin2016} and
 selective photo-dissociation \citep{Miotello2016}. Here we show an even more fundamental problem of 
using CO observations to determine the disk gas mass, namely, the diversity of vertical temperature structure among disks, and its impact on CO vertical gas abundance profiles in individual disks. 

Detailed modeling of disk SEDs, as shown above, reveal a range of thicknesses for VIRaM layers.
The vertical temperature structure regulates the vertical distribution of CO gas and ice 
and hence the amount of gas-phase CO which can be traced through (optically 
thin) CO isotopologue observations. We demonstrate the magnitude of this effect in Figure.~\ref{fig:comodels}, where we model the distribution of CO in two DIAD models with the same 
midplane temperatures, but with different vertical temperature profiles 
(cf. Figure~\ref{fig:n2hpsnow}). 
We assume that [$^{12}$CO]/[H$_2$]=10$^{-4}$ where $T>T_{\rm CO}$ and that it is reduced by two orders of magnitude where $T<T_{\rm CO}$, and make predictions for two cases:  T$_{\rm CO}$=20 and 28 K, corresponding to CO freeze-out on water-poor and water-rich ices, respectively. Figure~\ref{fig:comodels} shows that for each assumed CO freeze-out temperature, disks with and without a thick VIRaM layer present different CO column density radial profiles, both in terms of shapes and absolute column densities. The latter varies by factors of 2--5 between the two disk models. 
In the model with a thick VIRaM layer, 
20\% (T$_{\rm CO}$=28 K) 
to 40\% (T$_{\rm CO}$=20 K) of the available CO is in the gas-phase within 200 AU. By contrast, in the model without such a thick VIRaM layer 90\% of the available CO is in the gas interior to 200~AU, regardless of CO freeze-out temperature. Note the subtle drop of the CO column density in this kind of model. 
In summary, in a disk with a set midplane temperature profile, 20--90\% of the total CO resides in the gas-phase. Without detailed disk models it is therefore challenging to accurately account for CO freeze-out when using CO gas lines to estimate disk gas masses.
To develop CO emission lines as accurate probes of disk gas mass instead requires high resolution observations of both CO isotopologues and \nnhp{}, which together can be used to constrain the disk vertical structure. 

\section{Summary}
We present high angular resolution \nnhp{} $3-2$ observations of 6 single-disk systems. These observations show that \nnhp{} probes both the CO and N$_2$ snow surfaces. Our findings are as follows:
\begin{itemize}
    \item We find two distinctive emission morphologies in the \nnhp{} $3-2$ emission: either in a bright, narrow ring surrounded by extended tenuous emission (LkCa~15, GM~Aur, DM~Tau) or in a broad ring for most of the emission (V4046~Sgr, AS~209, IM~Lup).
    \item The bright, narrow ring pattern can be explained by \nnhp emission tracing vertical snow surfaces of CO and N$_2$  in disks with a thick VIRaM layer. In these disks, we use the inner and outer edges of the bright \nnhp{} ring to constrain the first set of CO and N$_2$ snowline pairs in disks. 
    \item Broad \nnhp{} rings are found in disks with a thin VIRaM layer, where the N$_2$ snowline cannot be constrained and only upper limits of the CO snowline locations can be obtained. 
    \item In disks where both CO and N$_2$ snowlines are located, we can determine the snowline temperatures based on the temperature structures of their respective disk models. The CO and N$_2$ snowline temperatures in the disks of LkCa~15 and GM~Aur are consistent with CO and N$_2$ freeze-out on moderately water-rich ice grains in an effectively static disk. However, those in the DM~Tau disk are considerably lower than the lowest expected value. 
\end{itemize}

Our observations and analysis show that the \nnhp{} imaging approach has tremendous potential to efficiently constrain the shape of CO and N$_2$ snow surfaces and the location of the corresponding snowlines. The results 
reveal a range of N$_2$ and CO snowline radii towards stars of similar spectra type, which demonstrate the need for empirically determined snowlines in disks.  

\acknowledgments We thank Ryan Loomis, Jane Huang, and Romane Le Gal for useful discussions. This paper makes use of ALMA data ADS/JAO.
ALMA\#2015.1.00678.S. ALMA is a partnership of ESO
(representing its member states), NSF (USA) and NINS
(Japan), together with NRC (Canada) and NSC and ASIAA
(Taiwan), in cooperation with the Republic of Chile. The Joint
ALMA Observatory is operated by ESO, AUI/NRAO and
NAOJ. The National Radio Astronomy Observatory is a
facility of the National Science Foundation operated under
cooperative agreement by Associated Universities, Inc. 
This paper utilizes the D'Alessio Irradiated Accretion Disk (DIAD) code. We wish to recognize the work of Paola D'Alessio, who passed away in 2013. Her legacy and pioneering work live on through her substantial contributions to the field. CCE acknowledges support from the National Science Foundation under Career grant AST-1455042. 

\facility{ALMA}
\software{CASA (McMullin et al. 2017), MIRIAD (Sault et al. 1995), DIAD (D'Alessio et al. 1998, 1999, 2001, 2005, 2006), RATRAN (Hogerheijde \& van der Tak 2000)}

\bibliographystyle{apj}



\appendix
\section{Channel Images}
\begin{figure*}[htbp]
\plotone{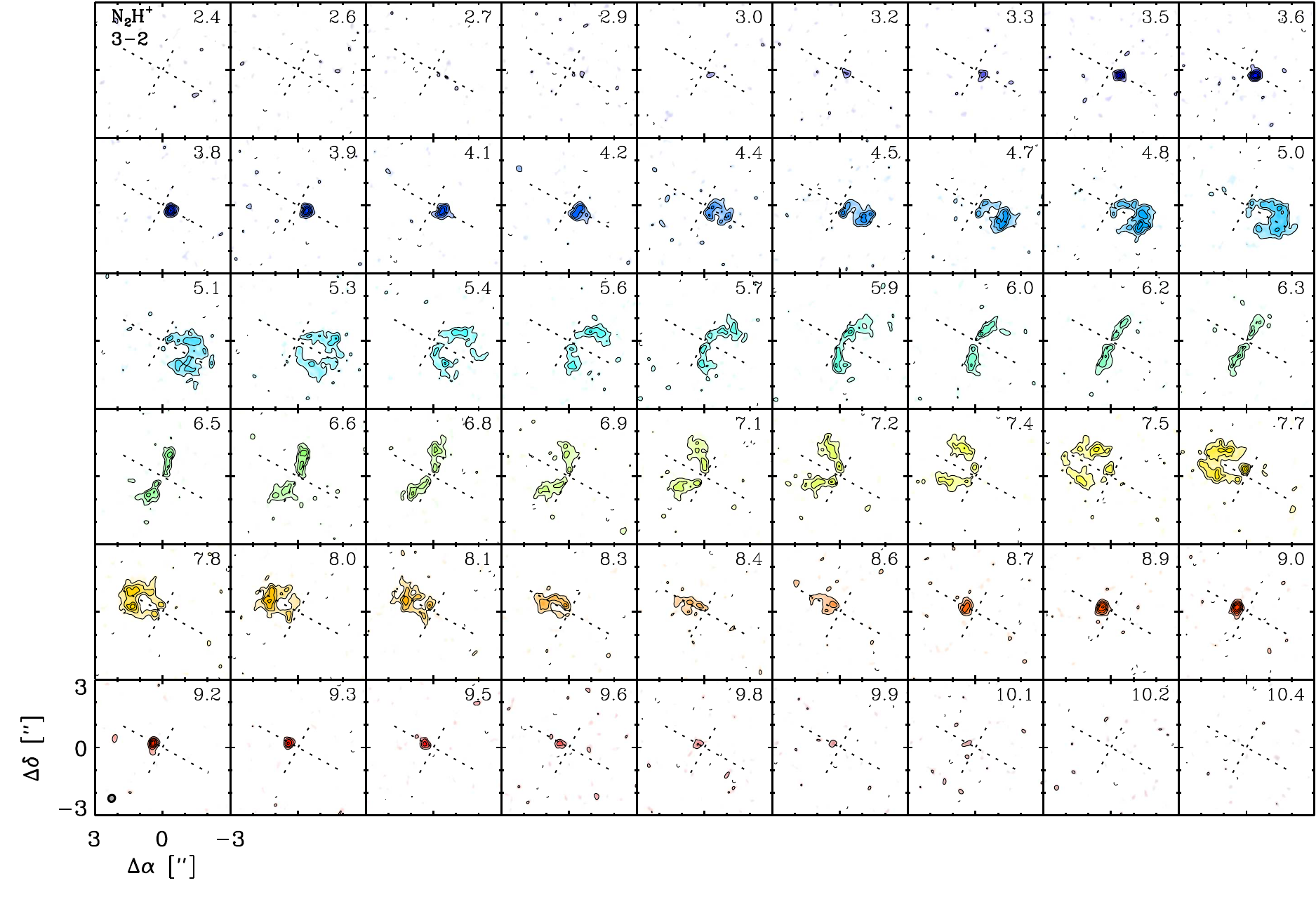}
\caption{\nnhp{} $3-2$ channel maps of the LkCa~15 disk. Contours are drawn at [3, 5, 7 ...]$\sigma$, where $\sigma$ is the channel rms listed in Table~\ref{tab:line}. Cross-hatches mark the stellar position and orientation of the disk position angle. The synthesized beam dimensions are drawn in the bottom left panel; LSR velocity (in km s$^{-1}$) are marked in the top right of each panel. The offset from the stellar position in arcseconds is marked on the axes in the lower left corner. 
\label{fig:lkca15channel}}
\end{figure*}

\begin{figure*}
\plotone{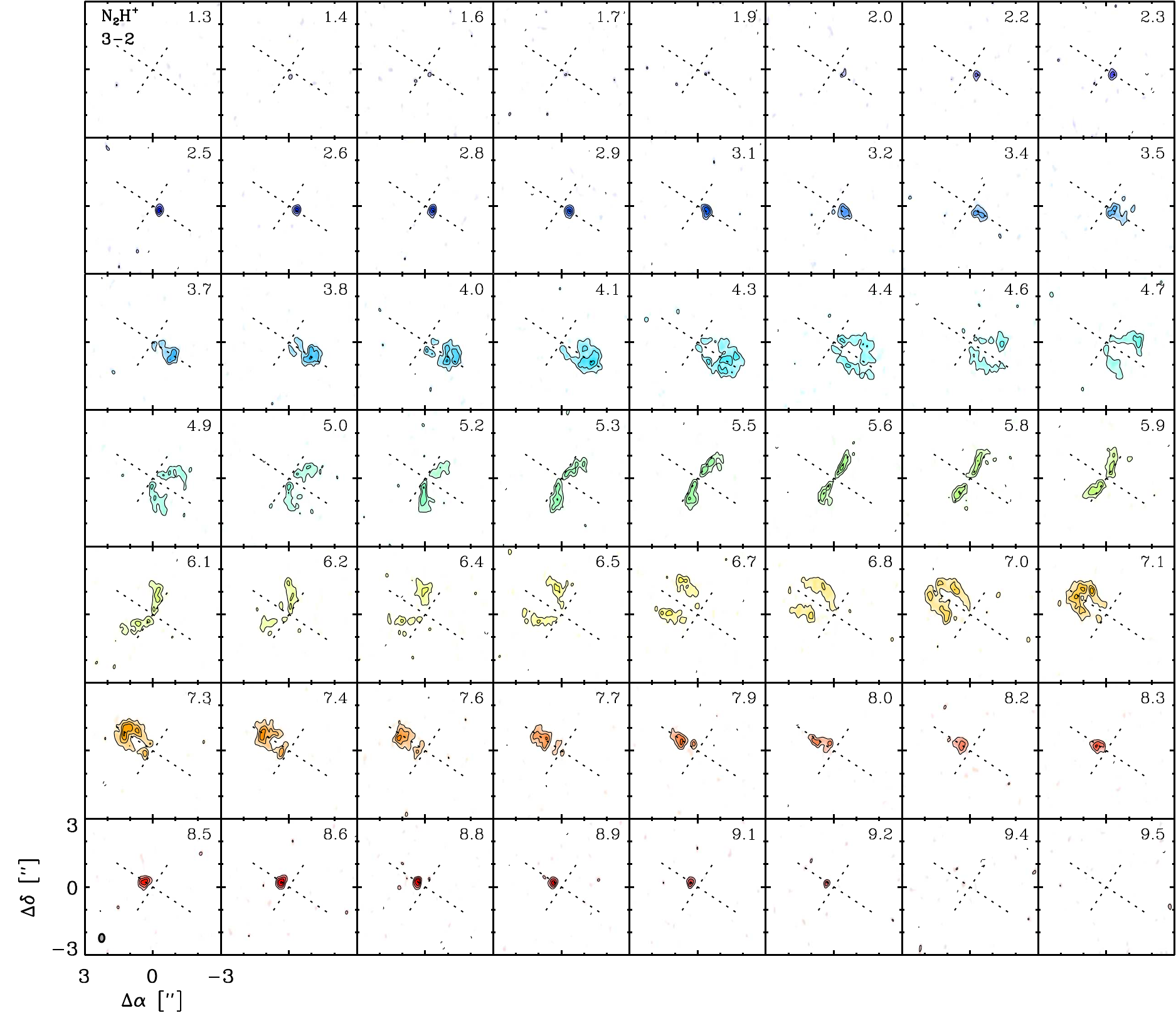}
\caption{\nnhp{} $3-2$ channel maps of the GM~Aur disk. \label{fig:gmaurchannel}}
\end{figure*}

\begin{figure*}
\plotone{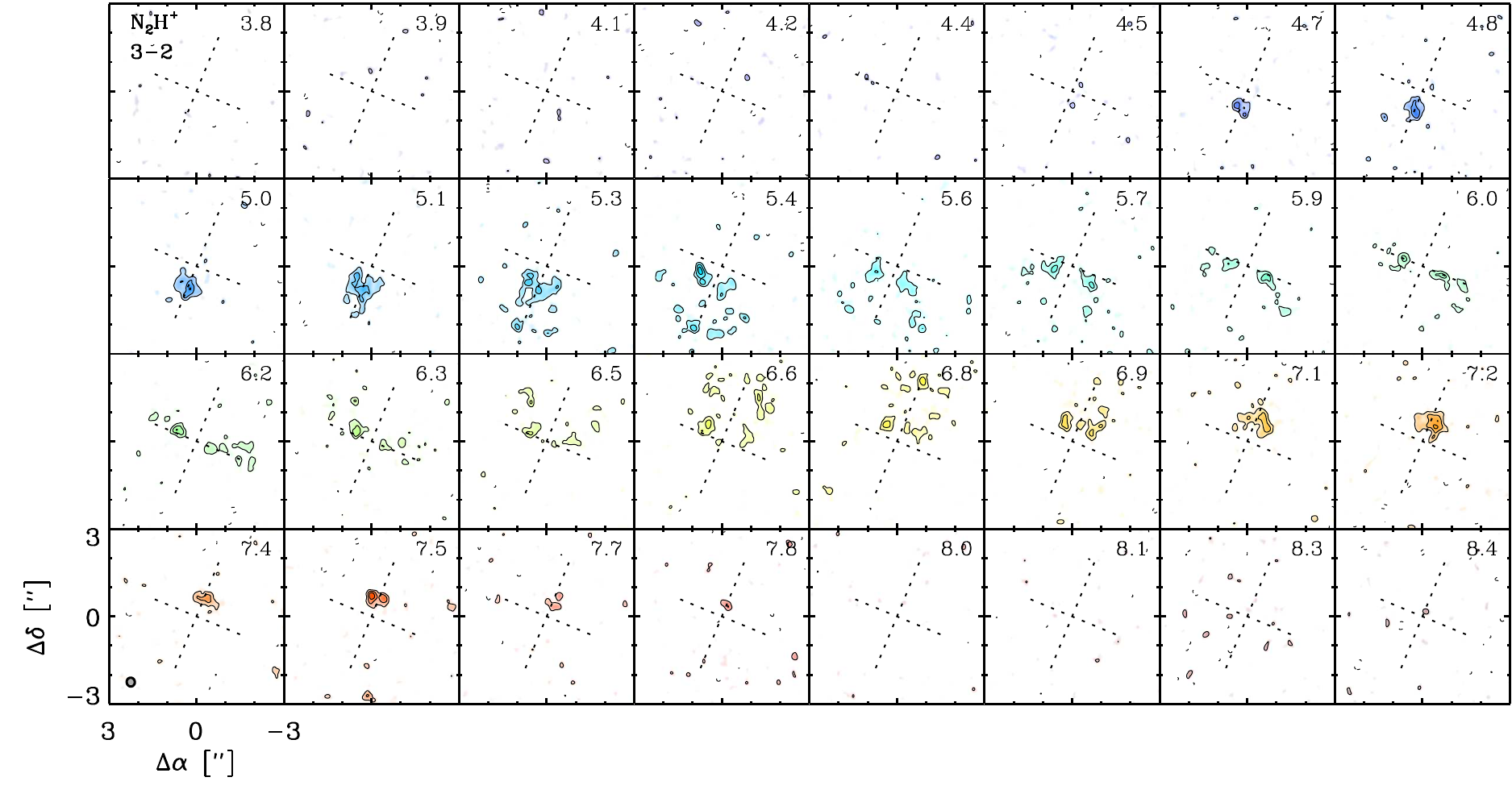}
\caption{\nnhp{} $3-2$ channel maps of the DM~Tau disk. \label{fig:dmtauchannel}}
\end{figure*}

\begin{figure*}
\plotone{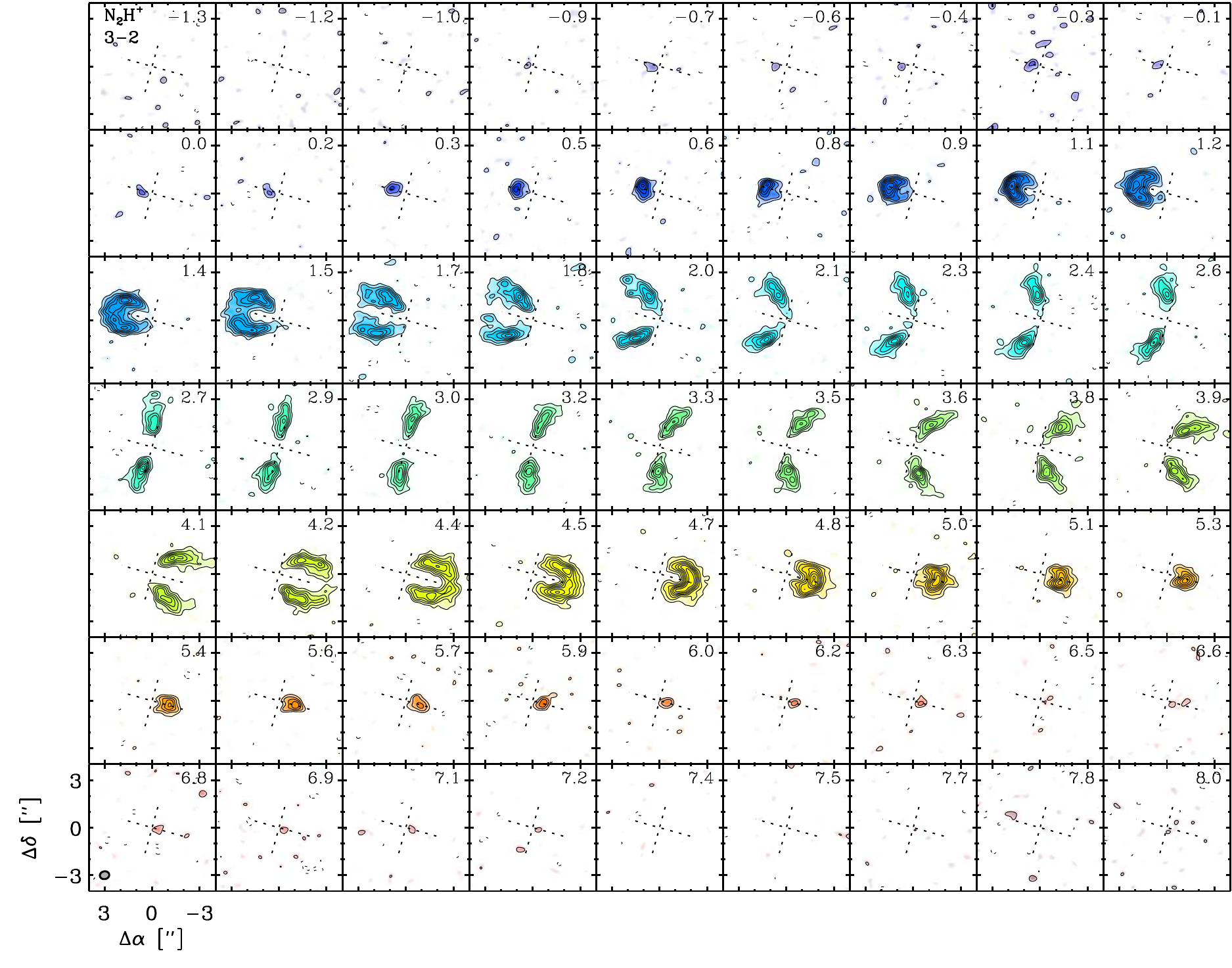}
\caption{\nnhp{} $3-2$ channel maps of the V4046~Sgr disk. \label{fig:v4046sgrchannel}}
\end{figure*}

\begin{figure*}
\plotone{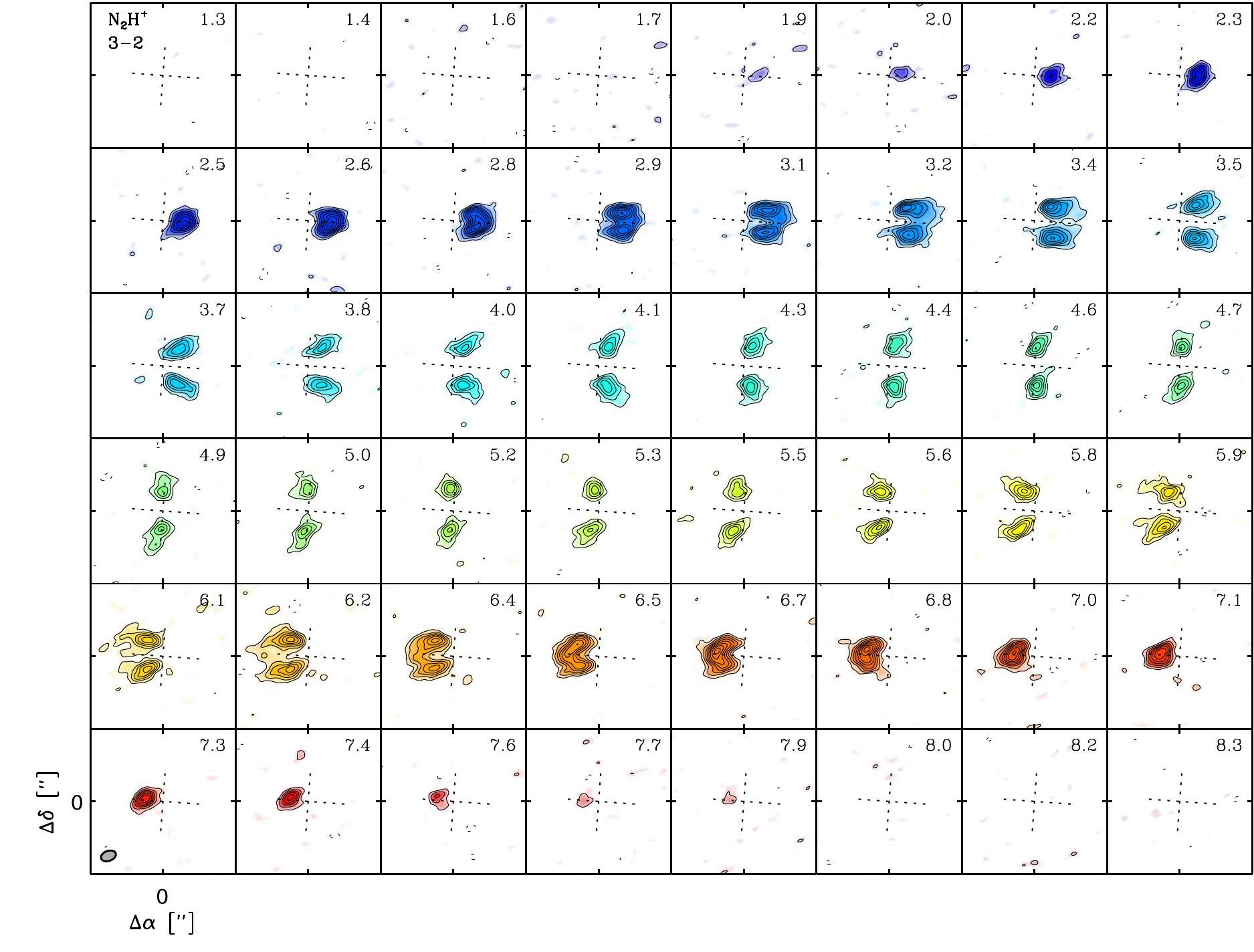}
\caption{\nnhp{} $3-2$ channel maps of the AS~209 disk. \label{fig:as209channel}}
\end{figure*}

\begin{figure*}
\plotone{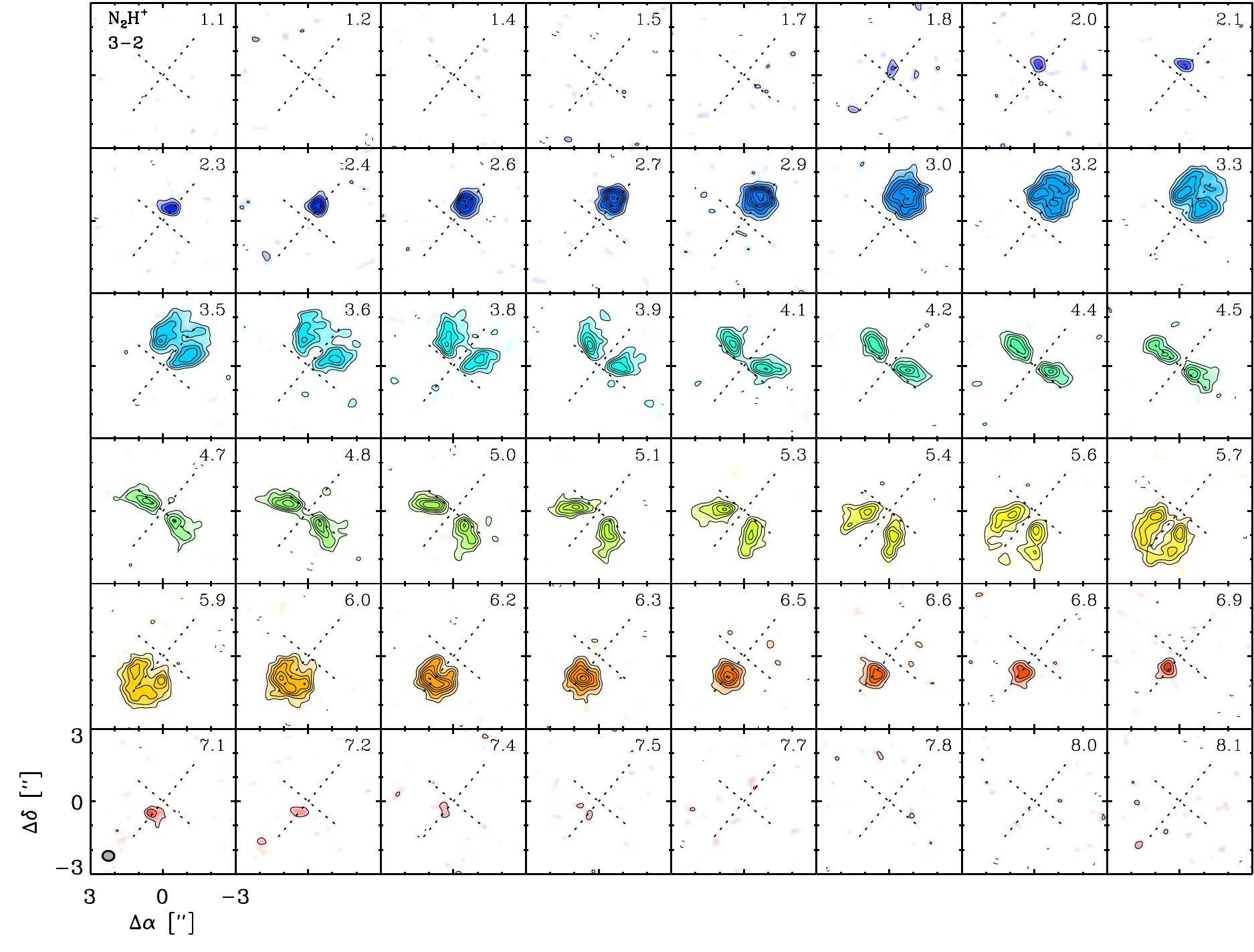}
\caption{\nnhp{} $3-2$ channel maps of the IM~Lup disk. \label{fig:imlupchannel}}
\end{figure*}

Figures~\ref{fig:lkca15channel}-~\ref{fig:imlupchannel} show channel maps for \nnhp{} $3-2$ emission toward the six disks. 

\section{Hyper-fine components of the \nnhp{} $3-2$ transition}
Table~\ref{tab:n2hphfs} lists the 29 \nnhp{} hyper-fine components 
of the $3-2$ transition and the Einstein A coefficients from the CDMS catalog. 
\begin{deluxetable*}{lc}
\tablecaption{Hyper-fine components of \nnhp{} $3-2$\label{tab:n2hphfs}}
\tablewidth{0pt}
\tablehead{
\colhead{Frequency (GHz)} & \colhead{A (10$^{-4}$ s$^{-1}$)}}  
\startdata
279.5094802 & 0.1212 \\
279.5098032 & 1.668 \\
279.5098546 & 1.572 \\
279.5098764 & 0.8648 \\
279.5102508 & 0.1304 \\
279.5111134 & 2.625 \\
279.5113207 & 8.162 \\
279.5113628 & 6.312 \\
279.5113923 & 0.6818 \\
279.5114098 & 1.706 \\
279.5114848 & 11.50 \\ 
279.5116237 & 0.3082 \\
279.5116627 & 10.77 \\
279.5117767 & 10.39 \\
279.5117843 & 11.82 \\
279.5117880 & 12.23 \\
279.5117885 & 12.84 \\
279.5118309 & 5.312 \\
279.5118379 & 13.52 \\
279.5121022 & 1.264 \\
279.5123006 & 1.244 \\
279.5138749 & 0.3675 \\
279.5139475 & 0.3759 \\
279.5140869 & 0.01079 \\
279.5142004 & 1.951 \\
279.5143188 & 1.713 \\
279.5143851 & 1.000 \\
279.5145717 & 0.2215 \\
279.5147106 & 0.5236 \\
\enddata
\end{deluxetable*}

\end{document}